# Eulerian-Lagrangian modelling of detonative combustion in two-phase gas−droplet mixtures with OpenFOAM: validations and verifications

Zhiwei Huang, Majie Zhao, Yong Xu, Guangze Li, Huangwei Zhang*

*Department of Mechanical Engineering, National University of Singapore, 9 Engineering Drive 1, Singapore, 117576, Republic of Singapore*

## Abstract

A hybrid Eulerian−Lagrangian solver RYrhoCentralFoam is developed based on OpenFOAM® to simulate detonative combustion in two-phase gas−liquid mixtures. For Eulerian gas phase, RYrhoCentralFoam enjoys second order of accuracy in time and space discretizations and is based on finite volume method on polyhedral cells. The following developments are made based on the standard compressible flow solver rhoCentralFoam in OpenFOAM®: (1) multi-component species transport, (2) detailed fuel chemistry for gas phase combustion, and (3) Lagrangian solver for gas-droplet two-phase flows and sub-models for liquid droplets. To extensively verify and validate the developments and implementations of the solver and models, a series of benchmark cases are studied, including non-reacting multi-component gaseous flows, purely gaseous detonations, and two-phase gas−droplet mixtures. The results show that the RYrhoCentralFoam solver can accurately predict the flow discontinuities (e.g. shock wave and expansion wave), molecular diffusion, auto-ignition and shock-induced ignition. Also, the RYrhoCentralFoam solver can accurately simulate gaseous detonation propagation for different fuels (e.g. hydrogen and methane), about propagation speed, detonation frontal structures and cell size. Sub-models related to the droplet phase are verified and/or validated against analytical and experimental data. It is also found that the RYrhoCentralFoam solver is able to capture the main quantities and features of the gas-droplet two-phase detonations, including detonation propagation speed, interphase interactions and detonation frontal structures. As our future work, RYrhoCentralFoam solver can also be extended for simulating two-phase detonations in dense droplet sprays.

**Keywords:** Shock wave; detonation wave; two-phase flow; liquid droplet; Eulerian−Lagrangian method; OpenFOAM

---

* Corresponding author. Tel.: +65 6516 2557; Fax: +65 6779 1459.
*E-mail address*: huangwei.zhang@nus.edu.sg.

## 1. Introduction

Understanding detonative combustion in different media is of great importance for engineering practice and hazard mitigation, e.g. in detonative combustion engines [1] and explosion [2–4]. Detonation wave runs supersonically (about 2,000 m/s) and is a combustion wave which couples the flame to a preceding shock wave. In crossing such a wave, the pressure and density considerably increase, which corresponds to unique solutions of the well-known Rankine-Hugoniot curves [5]. Detonative combustion in two-phase gas−liquid medium attracts increased interests in recent years, particularly due to the revived research thrust from aerosol or spray detonation propulsion exploiting liquid fuels [6].

Compared to detonation in homogeneous gas fuels, two-phase detonation introduces multi-facet complexities due to the addition of the dispersed phase in continuous phase. The droplet interacts with the surrounding gas, through mass, momentum, energy and species exchanges, which is expected to considerably change the chemico-physical properties of continuous gas phase where chemical reaction proceeds [7]. In turn, due to the spatially distinct gas fields caused by the detonation waves or shock/expansion waves, droplet may experience sharply evolving gas environment when it is dispersed and hence would demonstrate different dynamics (e.g. evaporation and heating [8]) from those in shockless flows. Therefore, to model detonative combustion in gas−droplet mixtures, it is of great significance to develop computationally accurate numerical algorithms to capture flow discontinuities and liquid droplet dynamics alike. Meanwhile, physically sound models are also necessitated to predict the correct droplet response to strong variations, temporally and spatially, of the surrounding gas, as well as two-way coupling between them.

There have been different numerical methods for simulations of two-phase detonations. For instance, Eulerian−Eulerian method is used by Wang and his co-workers [9] to simulate the droplet phase in two-phase detonations, while the space−time conservation element and solution element (CE/SE) schemes to capture the flow discontinuities. The Eulerian−Eulerian method is also used by Hayashi et al. [10] to model two-phase detonation and rotating detonation combustion. Although the computational cost of Eulerian−Eulerian method is low, however, individual droplet behaviors cannot be calculated, and only averaged quantities of the droplet phases are solved. In the meantime, the Eulerian droplet equations are only valid in the domain where droplets are statistically densely distributed. Otherwise, it may result in physical inconsistency and/or numerical instability. Conversely, Lagrangian tracking of individual particles enjoys numerous benefits, e.g. easy implementations and accurate descriptions of the instantaneous droplet locations and properties. It has been successfully applied for two-phase detonation simulations. For instance, Schwer et al. [11] use Eulerian−Lagrangian method with flux-corrected transport algorithms for droplet-laden detonations. Also, it is used by Zhang et al. [12] to simulate gas−solid two-phase detonations, together with CE/SE method as the shock-capturing scheme. More recently, it is employed with WENO (weighted essentially non-oscillatory) scheme for modelling the gas-droplet detonative flows by Ren et al. [13] and Watanabe et al. [14].

The open source package, OpenFOAM® [15], has proved to be a versatile and accurate code framework, which has been successfully used for modelling various fluid mechanics problems, including reacting compressible flows (e.g. by Huang et al. [16]) and multiphase flows (e.g. by Sitte et al. [17] and Huang et al. [18]). The existing density-based solver in OpenFOAM®, rhoCentralFoam, is deemed suitable for high-speed flows with shock and expansive waves [19]. It is based on finite volume discretization for polyhedral cells and enjoys second-order accurate for both spatial and temporal discretizations. In rhoCentralFoam, shock wave is accurately captured with the central-upwind Kurganov and Tadmor (KT) [20] or Kurganov, Noelle and Petrova (KNP) [21] scheme. Both schemes are computationally efficient, since complicated manipulations (e.g. characteristic decomposition or Jacobian calculation) are avoided for flux calculations on polyhedral cells. Greenshields et al. [19] validate this solver and found that rhoCentralFoam can accurately predict shock and expansion waves in different supersonic flows.

Recently, rhoCentralFoam is extended for modelling detonative combustion by Gutiérrez Marcantoni et al. [22], through incorporating multi-species transport and chemical reaction. They examine the capacities of rhoCentralFoam in predicting propagation of one-dimensional detonation waves in hydrogen/air mixtures, and the accuracies of their implementations for detonation modelling are validated [22]. Later, it is further used for capturing two-dimensional detonation propagation by them [23,24]. However, to the best of our knowledge, no efforts have been reported based on OpenFOAM® for two-phase gas−droplet detonations.

In this work, we aim to develop a high-fidelity numerical solver (termed as RYrhoCentralFoam hereafter) based on rhoCentralFoam in OpenFOAM® for simulating detonations in two-phase gas−droplet mixtures. Lagrangian method is used for tracking the liquid droplets. To this end, we make the following developments and implementations based on rhoCentralFoam: (1) multi-component transport, (2) detailed fuel chemistry for combustion, and (3) Lagrangian solver for two-phase flows and sub-models for liquid droplets. The foregoing developments and implementations are verified and validated in detail through well-chosen benchmark cases. They include non-reacting multi-component single-phase flows, purely gaseous detonations, and finally two-phase gas−droplet detonations. The rest of the manuscript is organized as below. The governing equations and computational method in RYrhoCentralFoam solver are presented in Section 2, followed by the validations and verifications in Section 3. The main findings are summarized in Section 4.

## 2. Governing equation and computational method

### *2.1 Governing equation for gas phase*

The governing equations of mass, momentum, energy, and species mass fractions, together with the ideal gas equation of state, are solved in RYrhoCentralFoam for compressible, multi-component, reacting flows. Due to the dilute droplet sprays considered in this work, the volume fraction effects from the dispersed phase on the gas phase are negligible [7]. Therefore, they are respectively written as

$$\frac{\partial \rho}{\partial t} + \nabla \cdot [\rho \mathbf{u}] = S_{mass}, \tag{1}$$

$$\frac{\partial(\rho\mathbf{u})}{\partial t} + \nabla\cdot[\mathbf{u}(\rho\mathbf{u})] + \nabla p + \nabla\cdot\mathbf{T} = \mathbf{S}_{mom}, \tag{2}$$

$$\frac{\partial(\rho \boldsymbol{E})}{\partial t} + \nabla\cdot[\mathbf{u}(\rho\boldsymbol{E})] + \nabla\cdot[\mathbf{u}p] - \nabla\cdot[\mathbf{T}\cdot\mathbf{u}] + \nabla\cdot\mathbf{j} = \dot{\omega}_T + S_{energy}, \tag{3}$$

$$\frac{\partial(\rho Y_m)}{\partial t} + \nabla\cdot[\mathbf{u}(\rho Y_m)] + \nabla\cdot\mathbf{s_m} = \dot{\omega}_m + S_{species,m}, (m = 1, \dots M-1), \tag{4}$$

$$p = \rho R T. \tag{5}$$

Here $t$ is time, $\nabla\cdot(\cdot)$ is divergence operator. $\rho$ is the density, $\mathbf{u}$ is the velocity vector, $T$ is the temperature, $p$ is the pressure and updated from the equation of state, i.e. Eq. (5). $Y_m$ is the mass fraction of $m$-th species, $M$ is the total species number. Only $(M-1)$ equations are solved in Eq. (4) and the mass fraction of the inert species (e.g. nitrogen) can be recovered from $\sum_{m=1}^{M} Y_m = 1$. $\boldsymbol{E}$ is the total energy, which is defined as $\boldsymbol{E} = e + |\mathbf{u}|^2/2$ with $e$ being the specific internal energy. $R$ in Eq. (5) is specific gas constant and is calculated from $R = R_u \sum_{m=1}^{M} Y_m MW_m^{-1}$. $MW_m$ is the molecular weight of $m$-th species and $R_u$ is universal gas constant. $\mathbf{T}$ in Eq. (2) is the viscous stress tensor, and modelled as

$$\mathbf{T} = 2\mu \mathrm{dev}(\mathbf{D}). \tag{6}$$

Here $\mu$ is dynamic viscosity, and is predicted with Sutherland's law, $\mu = A_s\sqrt{T}/(1 + T_S/T)$. Here $A_s = 1.67212 \times 10^{-6}$ kg/m $\cdot$ s $\cdot \sqrt{K}$ is the Sutherland coefficient, while $T_S = 170.672$ K is the Sutherland temperature. Moreover, $\mathbf{D} \equiv [\nabla\mathbf{u} + (\nabla\mathbf{u})^T]/2$ is deformation gradient tensor and its deviatoric component, i.e. $\mathrm{dev}(\mathbf{D})$ in Eq. (6), is defined as $\mathrm{dev}(\mathbf{D}) \equiv \mathbf{D} - \mathrm{tr}(\mathbf{D})\mathbf{I}/3$ with $\mathbf{I}$ being the unit tensor. In addition, $\mathbf{j}$ in Eq. (3) is the diffusive heat flux and can be represented by Fourier's law, i.e.

$$\mathbf{j} = -k\nabla T \tag{7}$$

with $k$ being the thermal conductivity, which is calculated using the Eucken approximation [25], $k = \mu C_v(1.32 + 1.37 \cdot R/C_v)$, where $C_v$ is the heat capacity at constant volume and derived from $C_v = C_p - R$. Here $C_p = \sum_{m=1}^{M} Y_m C_{p,m}$ is the heat capacity at constant pressure, and $C_{p,m}$ is estimated from JANAF polynomials [26].

In Eq. (4), $\mathbf{s_m} = -D\nabla(\rho Y_m)$ is the species mass flux. With the unity Lewis number assumption,

the mass diffusivity $D$ is calculated through the thermal conductivity as $D = k/\rho C_p$. $\dot{\omega}_m$ in Eq. (4) is the net production rate of $m$-th species due to chemical reactions and can be calculated from the reaction rate of each elementary reactions $\omega_{m,j}^o$, i.e.

$$\dot{\omega}_m = MW_m \sum_{j=1}^{N} \omega_{m,j}^o. \tag{8}$$

Here $N$ is the number of elementary reactions and $N > 1$ when multi-step or detailed chemical mechanism is considered. Here $\omega_{m,j}^o$ is calculated from

$$\omega_{m,j}^o = \left(v_{m,j}'' - v_{m,j}'\right)\left\{K_{fj} \prod_{m=1}^{M} [X_m]^{v_{m,j}'} - K_{rj} \prod_{m=1}^{M} [X_m]^{v_{m,j}''}\right\}. \tag{9}$$

$v_{m,j}''$ and $v_{m,j}'$ are the molar stoichiometric coefficients of $m$-th species in $j$-th reaction, respectively. $K_{fj}$ and $K_{rj}$ are the forward and reverse rates of $j$-th reaction, respectively. $[X_m]$ is molar concentration and calculated from $[X_m] = \rho Y_m / MW_m$. The combustion heat release, $\dot{\omega}_T$ in Eq. (3), is estimated as $\dot{\omega}_T = -\sum_{m=1}^{M} \dot{\omega}_m \Delta h_{f,m}^o$, with $\Delta h_{f,m}^o$ being the formation enthalpy of $m$-th species.

For gas−liquid two-phase flows, full coupling between the continuous gas phase (described by Eqs. 1-5) and dispersed liquid phase (described by Eqs. 10-12) is taken into consideration, in terms of the interphase exchanges of mass, momentum, energy and species. These respectively correspond to the source terms in the RHS of Eqs. (1)-(4), i.e. $S_{mass}$, $\mathbf{S}_{mom}$, $S_{energy}$ and $S_{species,m}$, and their equations are given in Eqs. (28)-(31). Nevertheless, if purely gaseous flows are studied, then these source terms are zero.

### *2.2 Governing equation for liquid phase*

The Lagrangian method is used in RYrhoCentralFoam to track the dispersed liquid phase which is composed of a large number of spherical droplets [27]. The interactions between the droplets are neglected, since only dilute sprays aim to be studied, in which the volume fraction of the dispersed droplet phase is generally less than 1‰ [7]. The governing equations of mass, momentum and energy for the individual droplets in the dispersed phase respectively read

$$\frac{dm_d}{dt} = \dot{m}_d, \tag{10}$$

$$\frac{d\mathbf{u}_d}{dt} = \frac{\mathbf{F}_d}{m_d}, \tag{11}$$

$$c_{p,d}\frac{dT_d}{dt} = \frac{\dot{Q}_c+\dot{Q}_{lat}}{m_d}, \tag{12}$$

where $m_d$ is the mass of a single droplet and can be calculated as $m_d = \pi\rho_d d_d^3/6$ for spherical droplets ($\rho_d$ and $d_d$ are the material density and diameter of a single droplet, respectively). $\mathbf{u}_d$ is the droplet velocity vector, $c_{p,d}$ is the droplet heat capacity, and $T_d$ is the droplet temperature. Both material density $\rho_d$ and heat capacity $c_{p,d}$ of the droplet are functions of droplet temperature $T_d$ [28], i.e.

$$\rho_d(T_d) = \frac{a_1}{a_2^{1+(1-T_d/a_3)^{a_4}}}, \tag{13}$$

$$c_{p,d}(T_d) = \frac{b_1^2}{\tau} + b_2 - \tau\left\{2.0b_1b_3 + \tau\left\{b_1b_4 + \tau\left[\frac{1}{3}b_3^2 + \tau\left(\frac{1}{2}b_3b_4 + \frac{1}{5}\tau b_4^2\right)\right]\right\}\right\}, \tag{14}$$

where $a_i$ and $b_i$ denote the species-specific constants and can be found from Ref. [28]. In Eq. (14), $\tau = 1.0 - min(T_d, T_c)/T_c$, where $T_c$ is the critical temperature (i.e. the temperature of a gas in its critical state, above which it cannot be liquefied by pressure alone) and $min\,(\cdot,\cdot)$ is the minimum function.

The evaporation rate, $\dot{m}_d$, in Eq. (10) is modelled through

$$\dot{m}_d = -\pi d_d Sh D_{ab} \rho_S ln(1 + B_M), \tag{15}$$

where $B_M$ is the Spalding mass transfer number estimated from Ref. [29]

$$B_M = \frac{Y_s - Y_g}{1 - Y_s}, \tag{16}$$

where $Y_s$ and $Y_g$ are respectively the vapor mass fractions at the droplet surface and in the ambient gas phase. $Y_s$ can be calculated as

$$Y_s = \frac{MW_d X_s}{MW_d X_s + MW_{ed}(1 - X_s)}, \tag{17}$$

where $MW_d$ is the molecular weight of the vapor, $MW_{ed}$ is the averaged molecular weight of the mixture excluding the vapor at the droplet surface, and $X_S$ is the mole fraction of the vapor at the droplet surface, which can be calculated using Raoult's Law

$$X_S = X_m \frac{p_{sat}}{p}, \tag{18}$$

in which $p_{sat}$ is the saturated pressure and is a function of droplet temperature $T_d$ [28], i.e.

$$p_{sat} = p \cdot exp\left(c_1 + \frac{c_2}{T_d} + c_3 lnT_d + c_4 T_d^{c_5}\right), \tag{19}$$

where $c_i$ are constants and can be found from Ref. [28]. The variation of $p_{sat}$ is expected to accurately reflect the liquid droplet evaporation in high-speed hot atmosphere, like in a shocked or detonated gas. In Eq. (18), $X_m$ is the molar fraction of the condensed species in the gas phase. Moreover, in Eq. (15), $\rho_S = p_S MW_m / RT_S$ is vapor density at the droplet surface, where $p_S = p \cdot exp\left(c_1 + \frac{c_2}{T_s} + c_3 lnT_s + c_4 T_s^{c_5}\right)$ is surface vapor pressure and $T_S = (T + 2T_d)/3$ is droplet surface temperature. $D_{ab}$ is the vapor mass diffusivity in the gaseous mixture, and modelled as [30]

$$D_{ab} = 3.6059 \times 10^{-3} \cdot (1.8T_s)^{1.75} \cdot \frac{\alpha}{p_s \beta}, \tag{20}$$

where $\alpha$ and $\beta$ are the constants related to specific species [30].

The Sherwood number $Sh$ in Eq. (15) is modelled as [31]

$$Sh = 2.0 + 0.6 Re_d^{1/2} Sc^{1/3}, \tag{21}$$

where $Sc$ is the Schmidt number of the gas phase. The droplet Reynolds number in Eq. (21), $Re_d$, is defined based on the velocity difference between two phases, i.e.

$$Re_d \equiv \frac{\rho_d d_d |\mathbf{u}_d - \mathbf{u}|}{\mu}. \tag{22}$$

In Eq. (11), only the Stokes drag is taken into consideration, and modeled as (assuming spherical droplets) [32]

$$\mathbf{F}_d = \frac{18\mu}{\rho_d d_d^2} \frac{C_d Re_d}{24} m_d (\mathbf{u} - \mathbf{u}_d). \tag{23}$$

The drag coefficient in Eq. (23), $C_d$, is estimated as [32]

$$C_d = \begin{cases} 0.424, & Re_d > 1000, \\ \frac{24}{Re_d}\left(1 + \frac{1}{6} Re_d^{2/3}\right), & Re_d < 1000. \end{cases} \tag{24}$$

The convective heat transfer rate $\dot{Q}_c$ in Eq. (12) is calculated by

$$\dot{Q}_c = h_c A_d (T - T_d), \tag{25}$$

where $A_d$ is the surface area of a single droplet. $h_c$ is the convective heat transfer coefficient, and computed from Nusselt number using the correlation by Ranz and Marshall [31]

$$Nu = \frac{h_c d_d}{k} = 2.0 + 0.6 Re_d^{1/2} Pr^{1/3}, \quad (26)$$

where $Pr$ is the gas Prandtl number. In addition, the latent heat of vaporization, $\dot{Q}_{lat}$ in Eq. (12), is

$$\dot{Q}_{lat} = -\dot{m}_d h(T_d), \quad (27)$$

where $h(T_d)$ is the vapor enthalpy at the droplet temperature $T_d$.

The two-way coupling terms, $S_{mass}$, $\mathbf{S}_{mom}$, $S_{energy}$ and $S_{species,m}$ in Eqs. (1)-(4), can be estimated based on the droplets in individual CFD cells, which read ($V_c$ is the cell volume, $N_d$ is the droplet number in the cell)

$$S_{mass} = -\frac{1}{V_c}\sum_1^{N_d} \dot{m}_d, \quad (28)$$

$$\mathbf{S}_{mom} = -\frac{1}{V_c}\sum_1^{N_d} \mathbf{F}_d, \quad (29)$$

$$S_{energy} = -\frac{1}{V_c}\sum_1^{N_d} (\dot{Q}_c + \dot{Q}_{lat}), \quad (30)$$

$$S_{species,m} = \begin{cases} S_{mass} & for\ condensed\ species, \\ 0 & for\ other\ species. \end{cases} \quad (31)$$

*2.3 Numerical implementation*

Finite volume method is used in RYrhoCentralFoam to discretize the Eulerian gas phase equations, i.e. Eqs. (1)-(4), over unstructured and arbitrary polyhedral cells [19]. Second-order backward scheme is employed for temporal discretization. The diffusion fluxes are calculated with second-order central differencing schemes. For the convection terms, the second-order semi-discrete and non-staggered KNP [21] scheme is used. The Gauss's divergence theorem can be written over a control volume $V$, i.e.

$$\int_V \nabla \cdot [\mathbf{u}\boldsymbol{\Psi}] dV = \int_S [\mathbf{u}\boldsymbol{\Psi}] dS \approx \sum_f \phi_f \Psi_f. \quad (32)$$

Here $\boldsymbol{\Psi}$ is a generic variable, representing $\rho$, $\rho\mathbf{u}$, $\rho\boldsymbol{E}$, or $p$. $S$ denotes the surfaces of a control volume. $\phi_f = \mathbf{S}_f \mathbf{u}_f$ is the volumetric flux across the surface $S$. $\sum_f$ means the summation over all the surfaces of the control volume $V$.

The sum of the flux in Eq. (32) can be written into three components [19,21]

$$\sum_f \phi_f \Psi_f = \sum_f \left[ \underbrace{\alpha\phi_{f+}\mathbf{\Psi}_{f+}}_{inward\ flux} + \underbrace{(1-\alpha)\phi_{f-}\mathbf{\Psi}_{f-}}_{outward\ flux} + \underbrace{\omega_f(\mathbf{\Psi}_{f+} - \mathbf{\Psi}_{f-})}_{weighted\ diffusion\ term} \right], \tag{33}$$

where α is weight factor. The first and second terms of the RHS of Eq. (33) denote the inward and outward fluxes, respectively. The third term is a diffusion term weighted by a volumetric flux $\omega_f$. For KNP scheme, biasness is introduced in the upwind direction, depending on the one-sided local speed of sound, leading to the central upwind characteristics of the KNP scheme. As such, α is calculated through $\psi_{f+}/(\psi_{f+} + \psi_{f-})$, and the volumetric fluxes $\psi_{f+}$ and $\psi_{f-}$ are calculated from the local speeds of propagation, i.e.

$$\psi_{f+} = max(c_{f+}|\mathbf{S}_f| + \phi_{f+}, c_{f-}|\mathbf{S}_f| + \phi_{f-}, 0). \tag{34}$$

$$\psi_{f-} = max(c_{f-}|\mathbf{S}_f| - \phi_{f+}, c_{f-}|\mathbf{S}_f| - \phi_{f-}, 0). \tag{35}$$

Here, $c_{f\mp} = \sqrt{\gamma R T_{f\mp}}$ are the sound speeds at the cell faces. To ensure the numerical stability, van Leer limiter [33] is used for correct numerical flux calculations with KNP scheme.

The individual liquid droplets are tracked with their Barycentric coordinates, parameterized by the topology (i.e. host cell the droplet lies in) and geometry (i.e. droplet location in the cell) (https://cfd.direct/openfoam/free-software/barycentric-tracking/). This approach is computationally efficient, and avoid the potential difficulties arising from mesh topology / quality and parallelization. The Lagrangian equations of liquid phase, i.e. Eqs. (10)-(12), are solved with first-order Euler method. The gas phase properties surrounding each droplet are interpolated to the droplet position from the closest cell centroids surrounding the droplet using linear interpolation.

It should be noted that the computational domains in OpenFOAM are always deemed three-dimensional (3D). For one-dimensional (1D) or two-dimensional (2D) cases studied in the following sections, the reduced direction(s) is discretized with only one cell and "*empty*" boundary condition (therefore without numerical flux calculations) is used in this direction. For two-phase flows, the dispersed droplets are essentially three-dimensional, and therefore the width of the 3D computational domain in the reduced direction(s) still plays a role in determining the droplet relevant quantities in each CFD cell (e.g. droplet number density or mass fraction).

## 3. Numerical validation and verification

### *3.1 Test cases of shock wave, species diffusion and chemical reaction*

#### *3.1.1 $O_2/N_2$ inert shock tube (Sod's shock tube problem)*

The Sod's shock tube problem [35] has been widely used for validating compressible flow solvers, e.g. in Refs. [36,37], to evaluate the dissipation of discontinuity-capturing schemes. The gas is $O_2/N_2$ mixture (i.e. 79%:21% by volume), instead of single-component gas, to examine the implementations of our species transport and thermal property calculations. For this case, we solve the 1D multicomponent Euler equations, neglecting the terms of viscous stress tensor, heat flux and species mass flux in Eqs. (2)-(4). The length of the computational domain is $L_x$ = 1 m, and it is discretized with 1,000 uniform cells. The initial conditions (at $t$ = 0, non-dimensional) of air correspond to the following Riemann problem:

$$(\rho, u, p) = \begin{cases} (1,0,1), & x \leq 0.5\, m \\ (0.125,0,0.1), & x > 0.5\, m \end{cases}. \tag{36}$$

The initial discontinuity would lead to right-propagating shock and contact discontinuity, as well as left-propagating rarefaction wave [35]. The CFL number is 0.02, which corresponds to a physical time step of about $10^{-8}$ s.

Figure 1 shows the comparison of numerical and analytical solutions for density, pressure, velocity, and speed of sound at $t$ = 0.007 s. Our numerical results are very close to the analytical solutions, and hence, our numerical solver shows good predictions for the Sod's shock tube problem with multicomponent air.

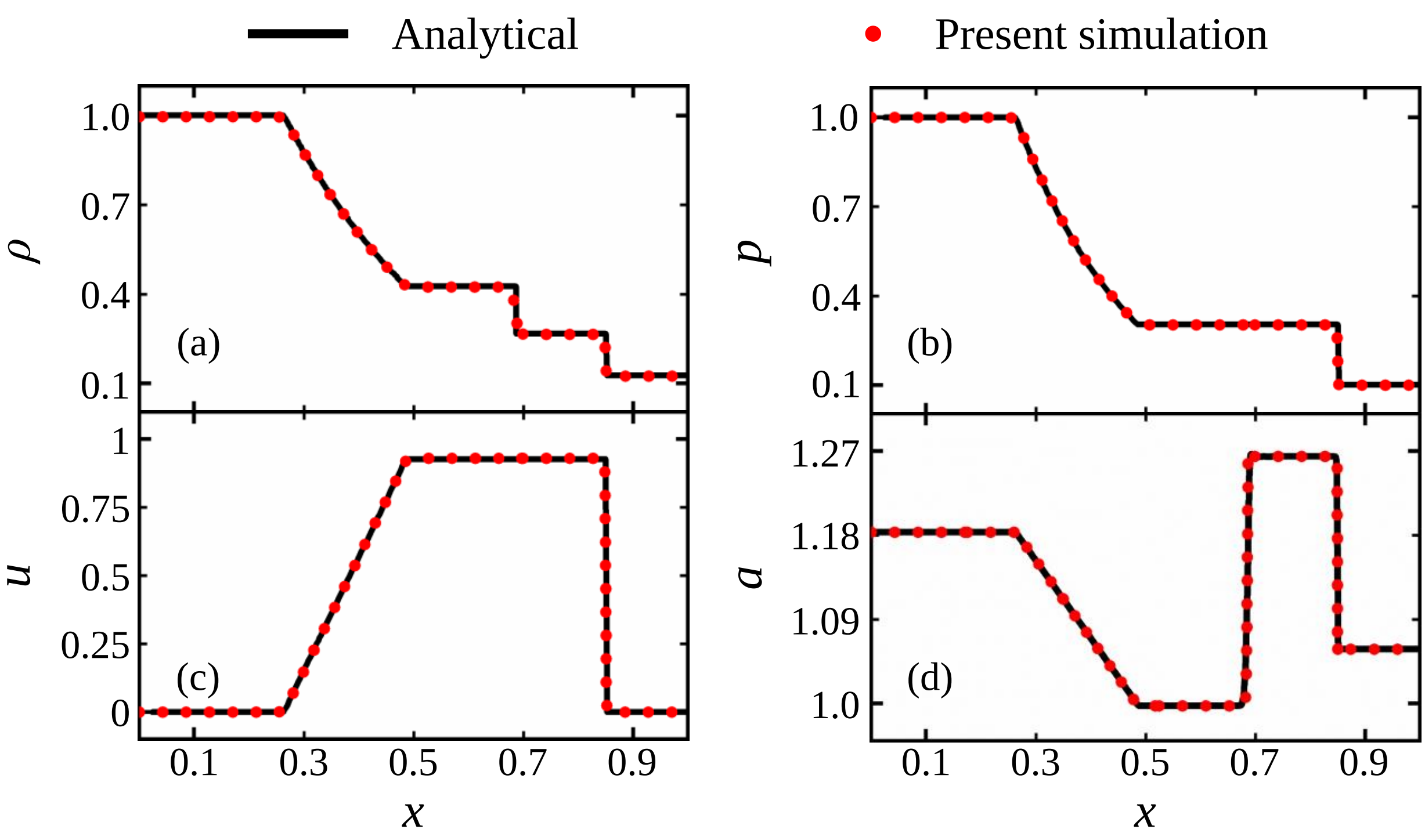

Figure 1. Comparisons between numerical and analytical solutions for the Sod's shock tube problem [35]: (a) density, (b) pressure, (c) velocity and (d) speed of sound at $t$ = 0.007 s.

### *3.1.2 Multicomponent ($H_2/O_2/Ar$) inert shock tube*

The shock tube with $H_2/O_2/Ar$ mixture is used to validate the accuracies in predictions of flow discontinuity and thermodynamics of the multicomponent mixture. It is a modified version of the Sod's shock tube problem [35], which is first investigated by Fedkiw et al. [38]. In this case, we solve the 1D Euler equations for an inert multicomponent mixture of $H_2/O_2/Ar$ with 2/1/7 by volume. The length of the computational domain is $L_x$ = 0.1 m, and it is discretized with 400 uniform cells. The CFL number is 0.02 (time step of about $10^{-8}$ s). The initial conditions (at $t$ = 0 μs) correspond to the following Riemann problem:

$$(T,u,p) = \begin{cases} (400\,K, 0\,m/s, 8{,}000\,Pa), & x \le 0.05\,m \\ (1{,}200\,K, 0\,m/s, 80{,}000\,Pa), & x > 0.05\,m \end{cases}. \tag{37}$$

Figure 2 shows the numerical solutions for density, temperature, velocity, and specific heat ratio at $t$ = 40 μs. The results show excellent agreement with those presented by Fedkiw et al. [38].

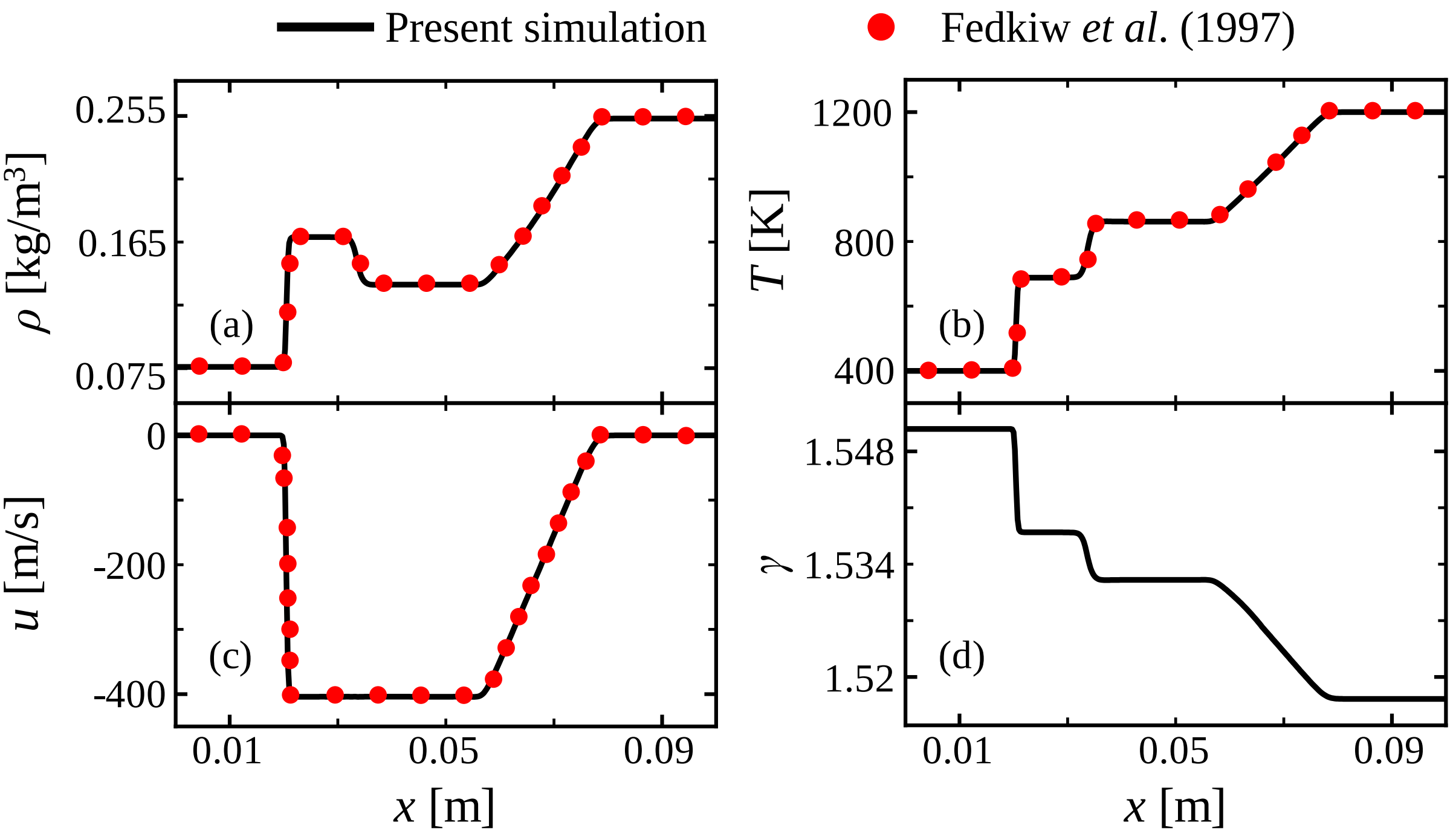

Figure 2. Comparisons between the numerical solutions for the multi-component inert shock tube problem [38]: (a) density, (b) temperature, (c) velocity and (d) specific heat ratio at $t = 40$ μs.

### *3.1.3 Multicomponent diffusion*

This case is used to evaluate the molecular diffusion terms in the Navier-Stokes equations, i.e. Eqs. (1)-(4). A multicomponent gas mixture consisting of $CH_4/O_2/H_2O/N_2$ in a 1D duct is simulated, which considers the simplified transport phenomena with unity Lewis number assumption. The duct is 0.05 m long and is discretized with uniform 200 cells. The CFL number is 0.02. Periodic boundary conditions are adopted at two sides of the computational domain. The initial pressure and velocity in the duct are 101,325 Pa and 0 m/s, respectively. The initial species mass fractions and temperature are given as below [39]

$$Y_m(x) = Y_{m,o} + \left(Y_{m,f} - Y_{m,o}\right) \cdot f(x), \tag{38}$$

$$T(x) = T_o + \left(T_f - T_o\right) \cdot f(x), \tag{39}$$

where $Y_m$ are the mass fractions of $CH_4$, $O_2$, $H_2O$, and $N_2$, $Y_{m,o}$ and $Y_{m,f}$ are their values at the oxidizer and fuel inlets, respectively. $T_o$ and $T_f$ are respectively the temperatures at the oxidizer and fuel inlets. The initial solution profile is defined by $f(x)$, which takes the following form

$$f(x) = 1 - exp\left[-\frac{(x-x_0)^2}{d^2}\right], \quad (40)$$

with $x_0 = 25 \times 10^{-2}$ m and $d = 2.5 \times 10^{-3}$ m. The inlet conditions for species and temperature are detailed in Table 1.

Table 1. Species mass fraction and temperature at the fuel and oxidizer inlets [39].

| Variables | $Y_{CH4}$ | $Y_{O2}$ | $Y_{H2O}$ | $Y_{N2}$ | $T$ [K] |
|---|---|---|---|---|---|
| Fuel | 0.214 | 0.195 | 0.0 | 0.591 | 320 |
| Oxidizer | 0.0 | 0.142 | 0.1 | 0.758 | 1,350 |

Figure 3 shows the profiles of $CH_4$ mass fraction and temperature at two different instants, which are compared with the multicomponent solutions by Vicquelin [39]. It is seen that the results in our simplified transport approximation (i.e. unity Lewis number) are in good agreement with those in the previous work [39]. Compared with the work of Martínez Ferrer et al. [37], our results also show good agreement with theirs using detailed transport model considering the Soret and Dufour effects.

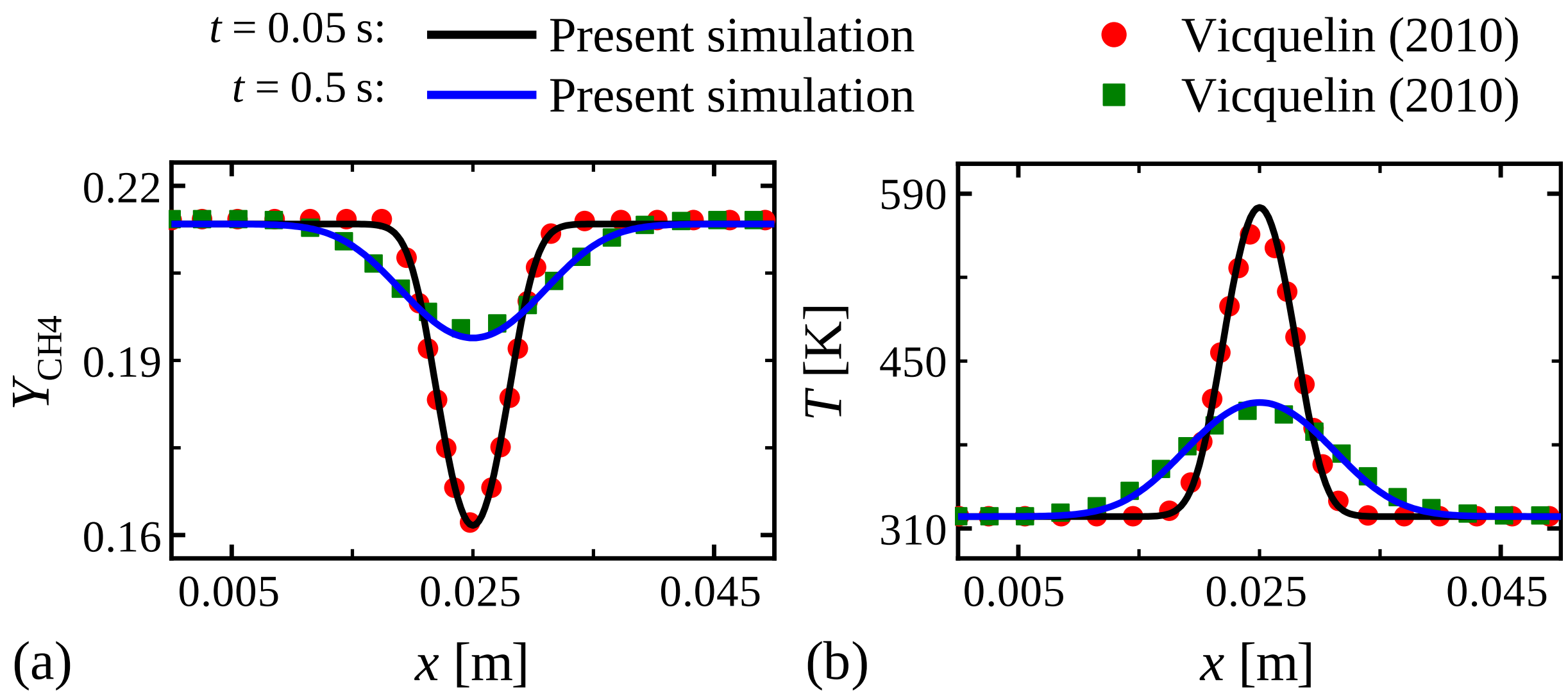

Figure 3. Comparisons between the numerical solutions for multicomponent diffusion [39]: (a) $CH_4$ mass fraction and (b) temperature at $t = 0.05$ s and 0.5 s.

*3.1.4 Perfectly stirred reactor*

This case focuses on the chemical source terms of the reactive Navier-Stokes equations, in terms of the reaction kinetic calculation and ODE (ordinary differential equation) solution method. For the constant volume auto-ignition of $H_2/O_2/N_2$ (2/1/7 by volume) mixture, the initial temperature and pressure are 1,000 K and 101,325 Pa, respectively. The governing equations of Eqs. (1)-(4) for this problem can be simplified to the following zero-dimensional equations for temperature and species mass fractions, i.e.

$$\frac{dE}{dt} = -\frac{1}{\rho}\sum_{m=1}^{M} \dot{\omega}_m \Delta h_{f,m}^{o}, \tag{41}$$

$$\frac{dY_m}{dt} = \frac{1}{\rho}\dot{\omega}_m. \tag{42}$$

Here $E$ only includes the sensible energy $e_s$, which is written as $e_s = \int_{T_0}^{T} C_v dT - R_u T_0/MW$.

A chemical mechanism of 9 species and 19 reactions for hydrogen [40] and a fixed time step of $10^{-6}$ s are used in our numerical simulations. A single cell with edges of 5 mm is used to mimic the constant volume autoignition, and "*empty*" boundary condition is applied for all the surfaces of the cell. Three different solvers for chemistry integration are tested, i.e. the Euler implicit solver (ODE solver of first-order accuracy), the Trapezoid solver (Trapezoidal ODE solver of second-order accuracy), and the rodas23 solver (low-stable, stiffly-accurate embedded Rosenbrock ODE solver of third-order accuracy) [41–43]. Other high-order accuracy ODE solvers available in OpenFOAM, e.g. RKCK45 (Cash-Karp Runge-Kutta ODE solver of 4/5th-order accuracy) [44], RKDP45 (Domand-Prince Runge-Kutta ODE solver of 4/5th-order accuracy) [45] and RKF45 (Runge-Kutta-Fehlberg ODE solver of 4/5th-order accuracy) [46] show similar accuracy with the 3rd-order rodas23 solver but with increased computational cost. Therefore, their results are not presented here.

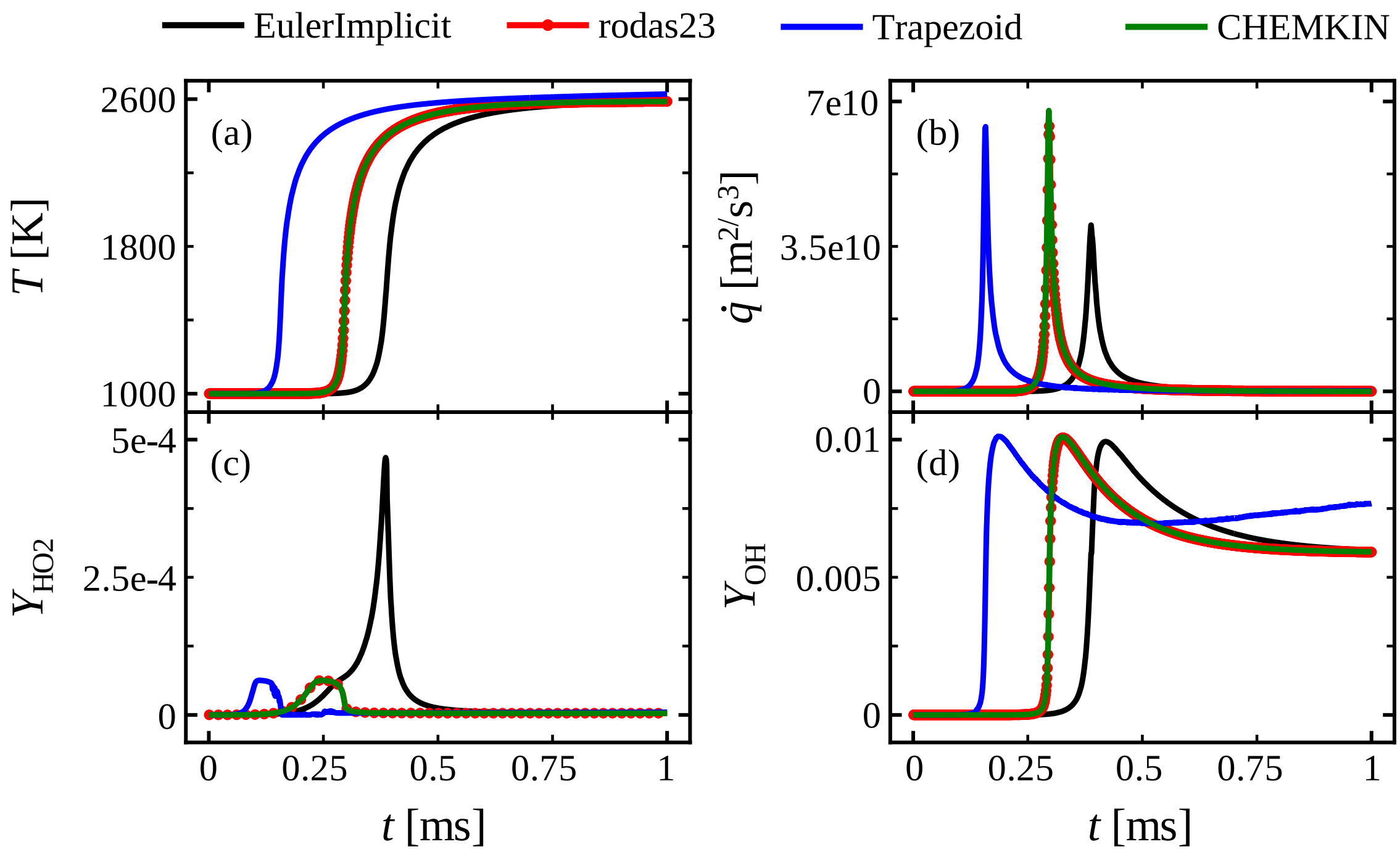

Figure 4. Comparisons between numerical solutions from OpenFOAM ODE solvers and CHEMKIN for auto-ignition of $H_2/O_2/N_2$ mixture in a perfectly stirred reactor: (a) temperature, (b) heat release rate, (c) $HO_2$ mass fraction, and (d) OH mass fraction.

Figure 4 shows the evolutions of temperature, heat release rate (divided by the constant density), $HO_2$ and OH mass fractions obtained from three different chemistry solvers. Solutions from the Perfectly Stirred Reactor (PSR) solver in the CHEMKIN library (in-house Fortran source code) [47] are also provided for comparisons, in which the same mechanism and time step are used. For this specific case, the results from rodas23 solver shows the excellent agreement with those from CHEMKIN. However, the Euler implicit and Trapezoid solvers have considerable discrepancies with the CHEMKIN solutions. Specifically, the Euler implicit solver over-predicts the ignition delay (of about 100 μs, estimated from the instant with maximum time derivative of temperature) and the peak value of $HO_2$ mass fraction (six times the value in the reference solution), but under-predicts the peak heat release rate. The Trapezoid solver under-predicts the ignition delay of about 160 μs but gives similar profiles for the shown variables (i.e. $T$, $\dot{q}$, $Y_{HO2}$, and $Y_{OH}$).

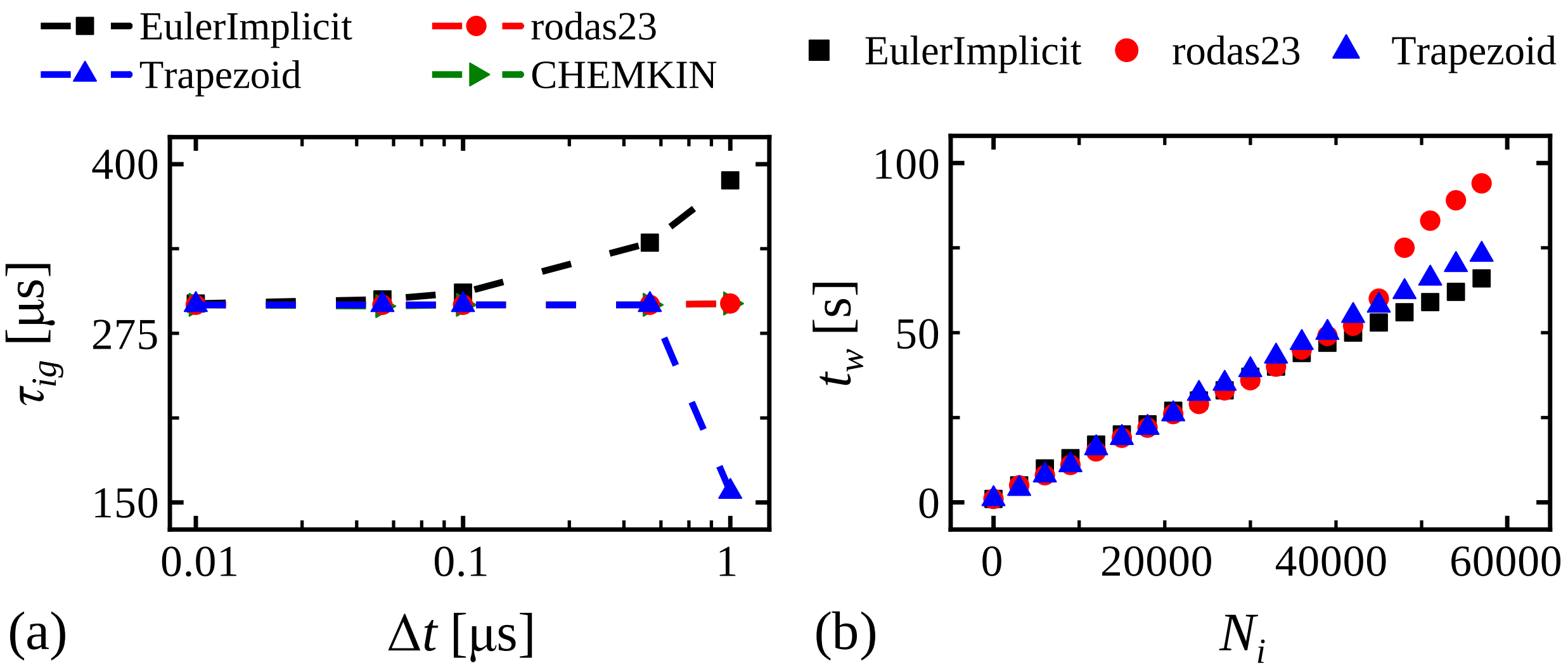

Figure 5. Comparisons of accuracy and efficiency for different chemistry solvers: (a) predicted ignition delay time versus computational time step, and (b) wall clock time versus number of time step.

However, the differences between the results of all the studied chemistry solvers are minimized when the time step is further decreased. Based on our numerical tests, they can yield close and accurate results (compared with the CHEMKIN solutions) when the time step is smaller than $5 \times 10^{-8}$ s for this case. Figure 5(a) shows the predicted ignition delay time ($\tau_{ig}$) versus the computational time step ($\Delta t$, logarithmic scale) for different chemistry calculation methods, i.e. Euler implicit, rodas23, and Trapezoid. It is seen that the Euler implicit solver over-predicts $\tau_{ig}$ when $\Delta t > 5 \times 10^{-8}$ s, and the Trapezoid solver under-predicts $\tau_{ig}$ when $\Delta t > 5 \times 10^{-4}$ s. When the time step is $\Delta t < 5 \times 10^{-8}$ s, the differences diminish. Figure 5(b) shows the wall clock time ($t_w$) required for the time steps ($N_i$) from the three chemistry solvers, at a fixed time step of $10^{-8}$ s. It is seen that the wall time of rodas23 solver is considerably longer than the other two chemistry solvers with increased iteration steps when noticeable combustion is initiated when ignition delay time is approached. Hence, the Euler implicit solver is most efficient among the three, and meanwhile has close accuracy to others when the time step is relatively small (e.g. lower than $5 \times 10^{-8}$ s as seen from Fig. 5a). Such small time steps are in reality frequently used in simulations of compressible flows to predict the highly unsteady aerodynamic phenomena, e.g. in Refs. [16,48,49], where the typical time step is as low as $10^{-9}$ s.

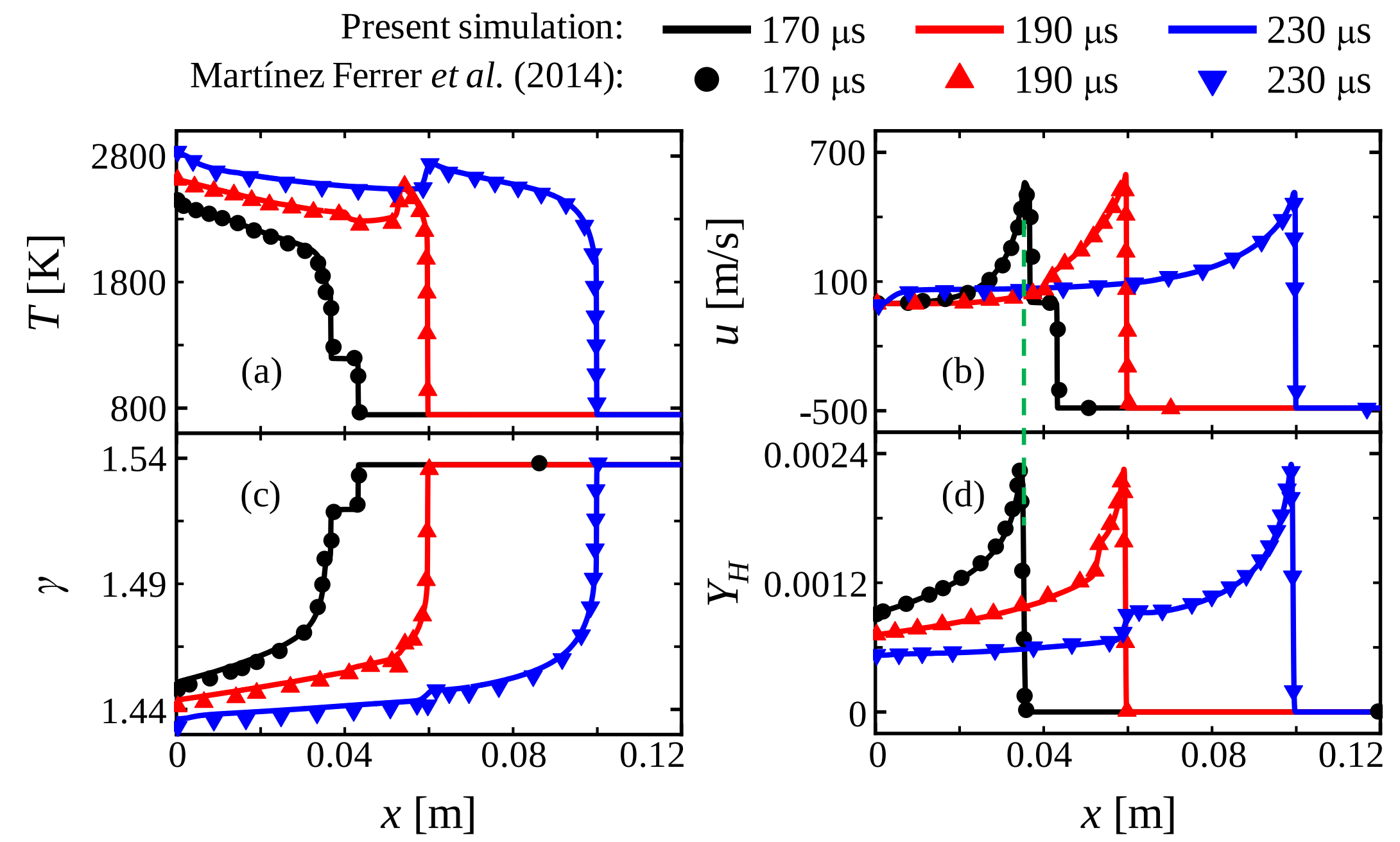

Figure 6. Comparisons between numerical solutions for the multi-component reactive shock tube problem [50]: (a) temperature, (b) velocity, (c) specific heat ratio and (d) H mass fraction at $t$ = 170 μs, 190 μs, and 230 μs.

### *3.1.5 Multicomponent ($H_2/O_2/Ar$) reactive shock tube*

The multicomponent ($H_2/O_2/Ar$) reactive shock tube is first investigated by Oran et al. [50], and has been widely used to evaluate the performance of numerical methods and implementations for compressible reactive flows, e.g. in Refs. [22,37,38,51]. Specifically, it can validate the accuracy of the solver in capturing the interactions between convection and reaction. A reactive mixture of $H_2/O_2/Ar$ with 2/1/7 by volume fills a semi-closed tube of 0.12 m long. The transport equations are 1D reactive multicomponent Euler equations. The 1D domain is discretized with 2,400 uniform cells. The CFL number is set as 0.02, corresponding to time step of about $10^{-8}$ s. The initial conditions (at $t$ = 0) correspond to:

$$(\rho, u, p) = \begin{cases} (0.072\ kg/m^3, 0\ m/s, 7{,}173\ Pa), & x \leq 0.06\ m \\ (0.18075\ kg/m^3, -487.34\ m/s, 35{,}594\ Pa), & x > 0.06\ m \end{cases}. \tag{43}$$

Solid wall conditions are set at the left boundary, while supersonic inlet condition is applied at the right boundary. A chemical mechanism of 9 species and 19 reactions for hydrogen [40] is used.

Figure 6 shows the distributions of temperature, velocity, specific heat ratio, and H mass fraction at three different instants. Results are compared with the simulation by Martínez Ferrer et al. [37], in which 7th-order WENO (Weighted Essentially Non-Oscillatory) scheme [52] and a mechanism of 9 species and 18 elementary reactions for hydrogen [53] are used. Generally, our results are quite close to their results, and also in excellent agreement with the results by Fedkiw et al. [38] (not shown in Fig. 6 for simplicity). Note that at $t$ = 170 μs, the reactive wave has not caught the reflected shock wave, as evident by the green dashed vertical line in Figs. 6(b) and 6(d). However, at $t$ = 190 and 230 μs, the reactive wave has merged with the reflected shock wave and detonative combustion occurs.

### *3.2 Single-phase detonation*

#### *3.2.1 One-dimensional hydrogen/air and methane/air detonation*

One-dimensional detonation propagation in premixed hydrogen/air and methane/air gas with different equivalence ratios are simulated. These tests aim to validate the accuracy of the RYrhoCentralFoam solver in predicting propagation speed of the detonation wave, which is a complex of preceding shock and auto-igniting reaction waves with an induction distance. The selected equivalence ratios are 0.5-4.0 for $H_2$ and 0.8-3.0 for $CH_4$, which lie in the detonability ranges of both mixtures as suggested by Glassman et al. [54]. The length of the 1D domain is 0.5 m, which is discretized by uniform cells of 0.02 mm for $H_2$ and 0.1 mm for $CH_4$, corresponding to more than 10 cells in respective Half-Reaction Length (HRL) of stoichiometric mixtures. This HRL is determined based on the distance between preceding shock front and the reaction front with maximum heat release in the ZND (Zel'dovich–Neumann–Döring) structures predicted by Shock & Detonation Toolbox [55], abbreviated as SD Toolbox hereafter. Mesh sensitivity analysis is performed based on finer cell sizes, and it is shown (results not presented here) that the above resolutions are sufficient in predictions of detonation propagation speed.

The initial temperature and pressure are $T_0$ = 300 K and $P_0$ = 1 atm, respectively. Moreover,

the left and right boundaries of the domain are assumed to be non-reflective. In OpenFOAM, the following equation is solved at the boundary

$$\frac{D\phi}{Dt} = \frac{\partial\phi}{\partial t} + \mathbf{u}\cdot\nabla\phi = 0, \quad (44)$$

where $\phi$ is a generic boundary variable and $\mathbf{u}$ is the velocity vector. Spurious wave reflections from the outlet boundary towards the interior domain is avoided with this non-reflective boundary condition.

Detailed mechanism (including 19 elementary reactions and 9 species) [56] is used for hydrogen combustion, which has been validated against the measured ignition delay at elevated pressures [57] and successfully applied for detonation modelling [58]. For methane, a skeletal mechanism with 35 reactions and 16 species [59] is used. The detonation is ignited by a hot spot (2 mm in width) at the left end of the domain, in which high temperature (2,000 K for $H_2$, 2,400 K for $CH_4$) and pressure (90 atm) are presumed. The conditions in the hot spot can successfully initiate the detonation waves, which quickly evolve into steady propagation at the speeds close to the Chapman–Jouguet (C–J) values.

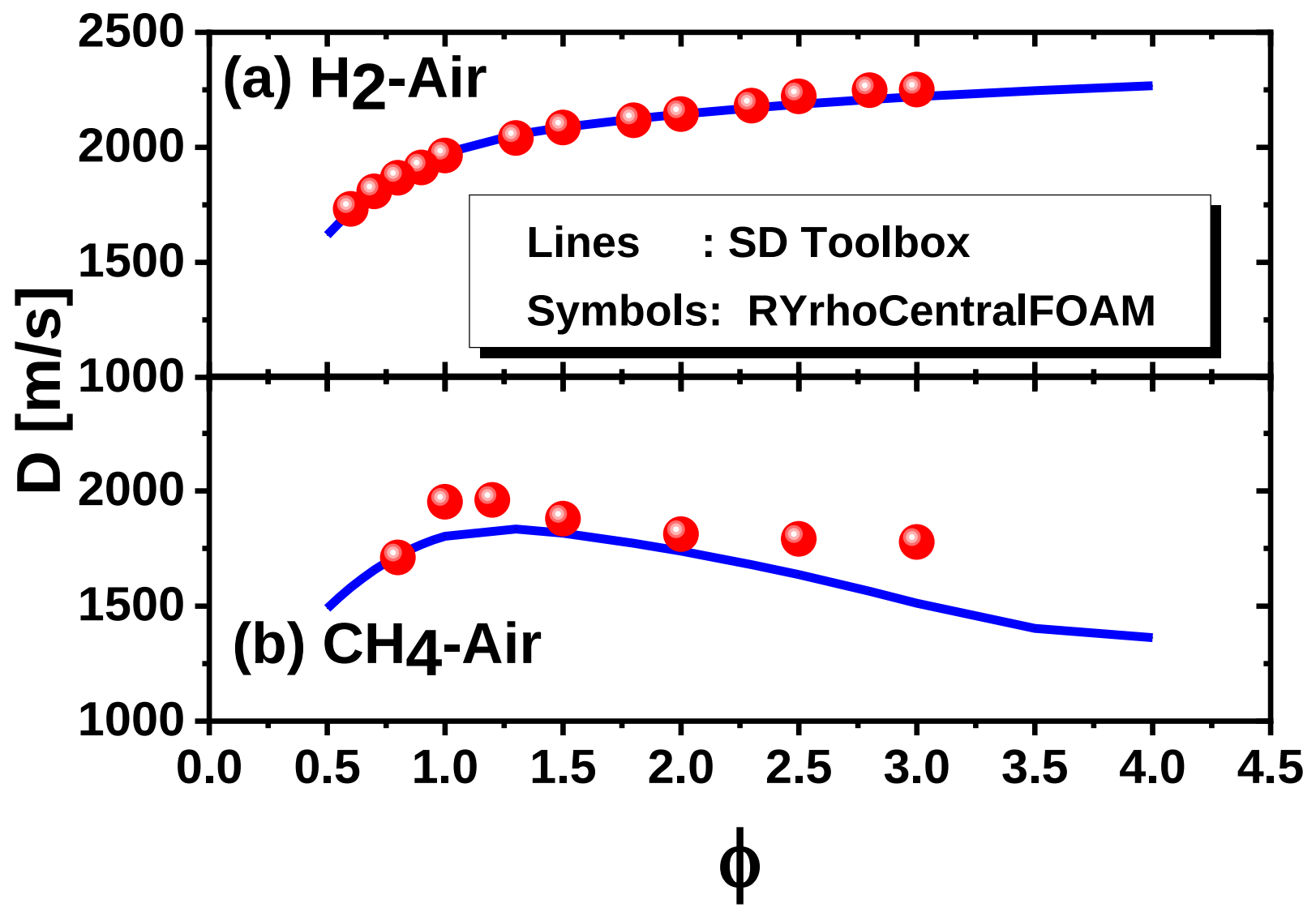

Figure 7. Detonation propagation speed versus equivalence ratio in (a) hydrogen/air and (b) methane/air mixtures. Symbols: RYrhoCentralFoam; lines: SD Toolbox [55].

Figures 7(a) and 7(b) respectively show the detonation propagation speed $D$ of hydrogen/air and methane/air mixtures as functions of equivalence ratios $\phi$. Here the speed is calculated based on the locations of the peak heat release over a fixed time interval, and $D$ in Fig. 7 is averaged from about 50 sampled speeds using the above method. For comparisons, we also add the C–J speeds predicted by SD Toolbox [55]. As demonstrated in Fig. 7(a), the detonation propagation speeds in $H_2$/air mixtures from RYrhoCentralFoam are in line with the results from SD Toolbox. For the methane/air results in Fig. 7(b), the agreements between the results from RYrhoCentralFoam and SD Toolbox are satisfactory for $0.8 < \phi < 2.5$. However, for fuel-richer case, e.g. $\phi = 3.0$ in Fig. 7(b), the propagation speed from RYrhoCentralFoam is slightly higher than that from SD Toolbox. It is likely due to the deposited hot spot which leads to some degree of overdrive for the travelling detonation waves at this peculiar equivalence ratio. In general, the results in Fig. 7 have confirmed the accuracy of RYrhoCentralFoam solver in calculating the propagation speed of 1D detonative combustion.

*3.2.2 Two-dimensional hydrogen/air detonation*

Two-dimensional detonation in premixed $H_2$/air mixtures is studied to examine the capacity and accuracy of the RYrhoCentralFoam solver to predict the cellular detonation front structure. Two equivalence ratios are considered, i.e. $\phi$ = 1.0 and 0.8. The computational domain is schematically demonstrated in Fig. 8. The length ($x$-direction) and width ($y$-direction) are 0.3 m and 0.01 m, respectively. The initial temperature and pressure in the domain are $T_0$ = 300 K and $P_0$ = 100 kPa, respectively. To reduce the computational cost, the domain is divided into three blocks (see Fig. 8, demarcated by dashed lines therein), with the individual resolutions varying from 0.1 mm in Block 1 to 0.01 mm in Block 3, which respectively lead to the total cells of 500,000, 1,280,000 and 5,000,000. Blocks 1 and 2 act as the driver section, whilst the discussion in this sub-section is based on the results from the finest Block 3 with approximately 20 cells in the HRL. The detonation is initiated through three vertically placed hot spots (100 atm and 2,000 K) at the left end to achieve the cellular detonative front within relatively short duration. The upper and lower boundaries in Fig. 8 are assumed to periodic, and the left and right sides are assumed to be non-reflective. The physical time step is $1\times10^{-9}$ s.

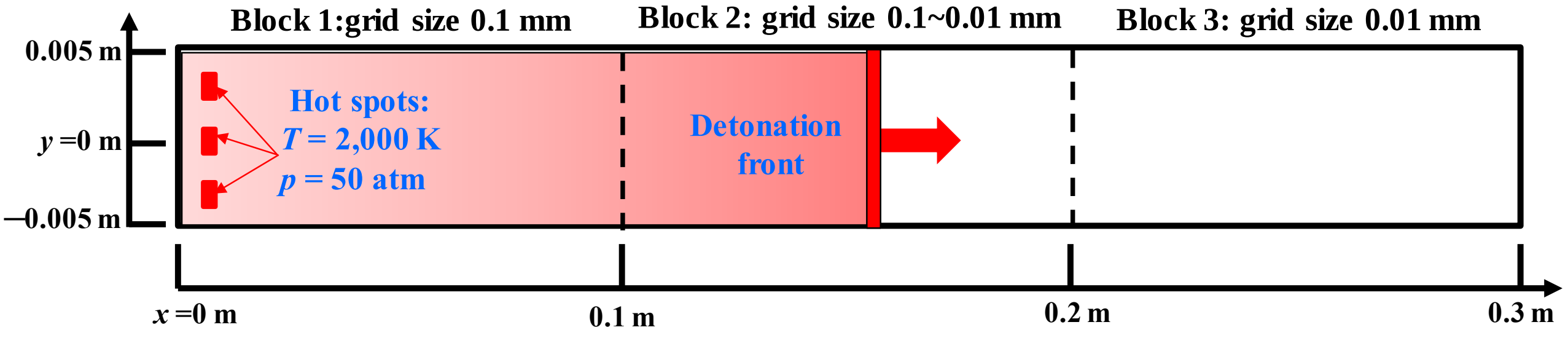

Figure 8. Schematic of computational domain.

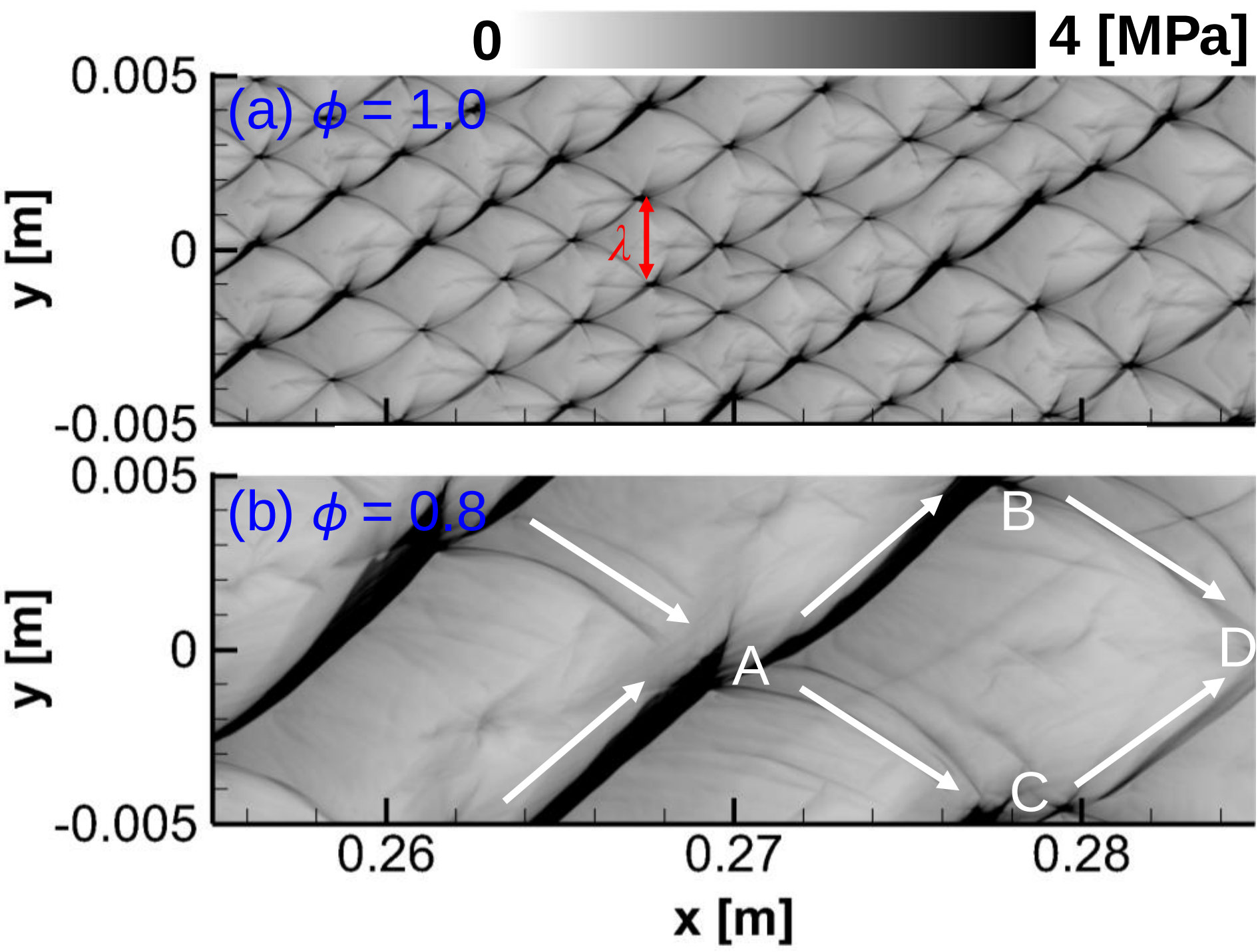

Figure 9. Peak pressure trajectory of hydrogen/air mixtures: (a) $\phi$ = 1.0 and (b) $\phi$ = 0.8. The white arrow indicates the movement direction of the triple points.

The history of maximum pressure during detonation propagation in two $H_2$/air mixtures is recorded in Fig. 9. The black or grey stripes in Fig. 9 essentially correspond to the trajectory of the triple points connecting the transverse wave, incident wave, Mach wave and shear layer [60]. The cell distributions are generally regular in both cases, and the cell sizes with $\phi$ = 1.0 in Fig. 9(a) are overall smaller than those with $\phi$ = 0.8 in Fig. 9(b). The averaged cell width $\lambda$ with both equivalence ratios are compared with the measured data [61,62] and theoretical estimations [63], as tabulated in Table 2.

It is found that with decreased $\phi$, λ increases, which agrees well with the measured and theoretical results. For $\phi$ = 1.0, our result is slightly under-predicted, whilst for $\phi$ = 0.8, our results show better agreement, particularly with the results by Ciccarelli et al. [62]. Therefore, the accuracies of RYrhoCentralFoam in calculations of detonation cell size are generally satisfactory.

Table 2. Cell widths of $H_2$/air mixtures with equivalence ratio 1.0 and 0.8

| | | Simulation | Experiment | | Theory |
|---|---|---|---|---|---|
| | | Present work | Guirao et al. [61] | Ciccarelli et al. [62] | Ng et al. [63] |
| Initial condition ($T_0$, $P_0$) | | 300 K, 100 kPa | 293 K, 101.3 kPa | 300 K,100 kPa | 300 K, 100 kPa |
| Cell width λ [mm] | $\phi$ = 1.0 | 2.8 | 15.1 | 8.19 ($\phi$ = 1.0233)† | 5.05 |
| | $\phi$ = 0.8 | 10.0 | 18.1 ($\phi$ = 0.7933) | 11.04 ($\phi$ = 0.79) | 7.08 |

† The equivalence ratio in the brackets indicate the actual value in the experiments.

### *3.3 Droplet phase sub-model*

#### *3.3.1 Droplet evaporation model*

Sub-models related to the droplet phase are validated and verified in this Section. Firstly, the droplet evaporation model detailed in Section 2.2 is verified through comparing the computational and analytical solutions about evaporation of a single water droplet in quiescent air. The square of droplet diameter can be obtained through integrating Eq. (10) assuming that evaporation rate coefficient $c_{evp}$ is constant, i.e. [64]

$$d_d^2(t) = d_d^2(t_0) - c_{evp}t \tag{45}$$

with $c_{evp} = 4\rho_s Sh D_{ab}\, ln(1 + B_M)/\rho_d$. Since droplet Reynolds number $Re_d$ << 1 in this case, $Sh$ ≈ 2.0 is valid (see Eq. 21). Three temperatures of surrounding gas are chosen, i.e. 400 K, 500 K, and 600 K, which result in different $c_{evp}$ of 9,102 μm²/s, 19,269 μm²/s, and 31,052 μm²/s, respectively. Note that here $c_{evp}$ is calculated *a posterior* based on the numerical results, and it is a time-averaged value, used to plot the analytical solution from Eq. (45). The air pressure is 1 atm, while the initial droplet temperature and diameter are 300 K and 100 μm, respectively.

Figure 10 shows the time history of droplet diameter squared at different air temperatures. Excellent agreement is found between the present simulations and the analytical solutions (i.e. Eq. 45) for all the three cases. Hence, the implementations of the evaporation model in RYrhoCentralFoam solver are correct.

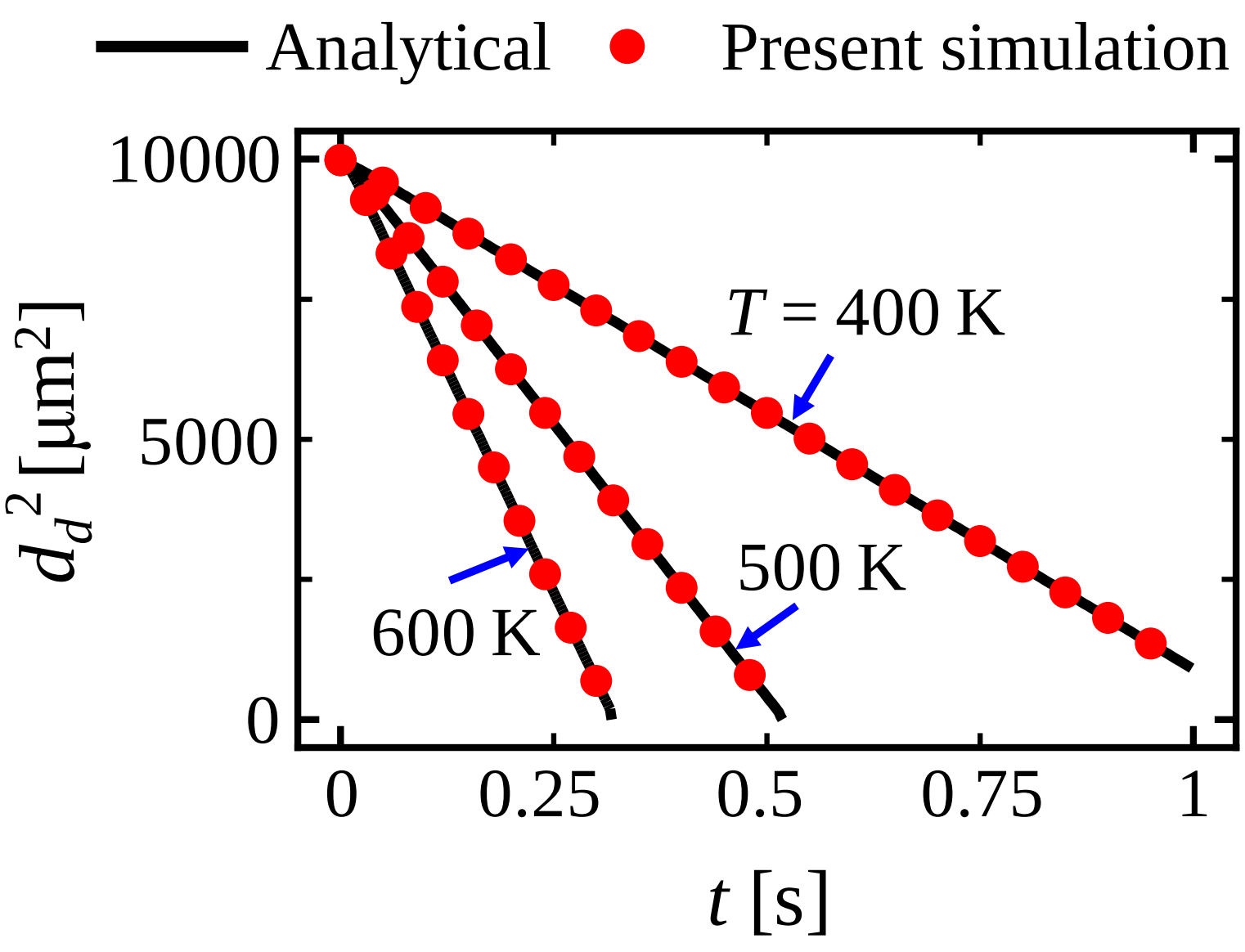

Figure 10. Time history of droplet diameter under different air temperatures

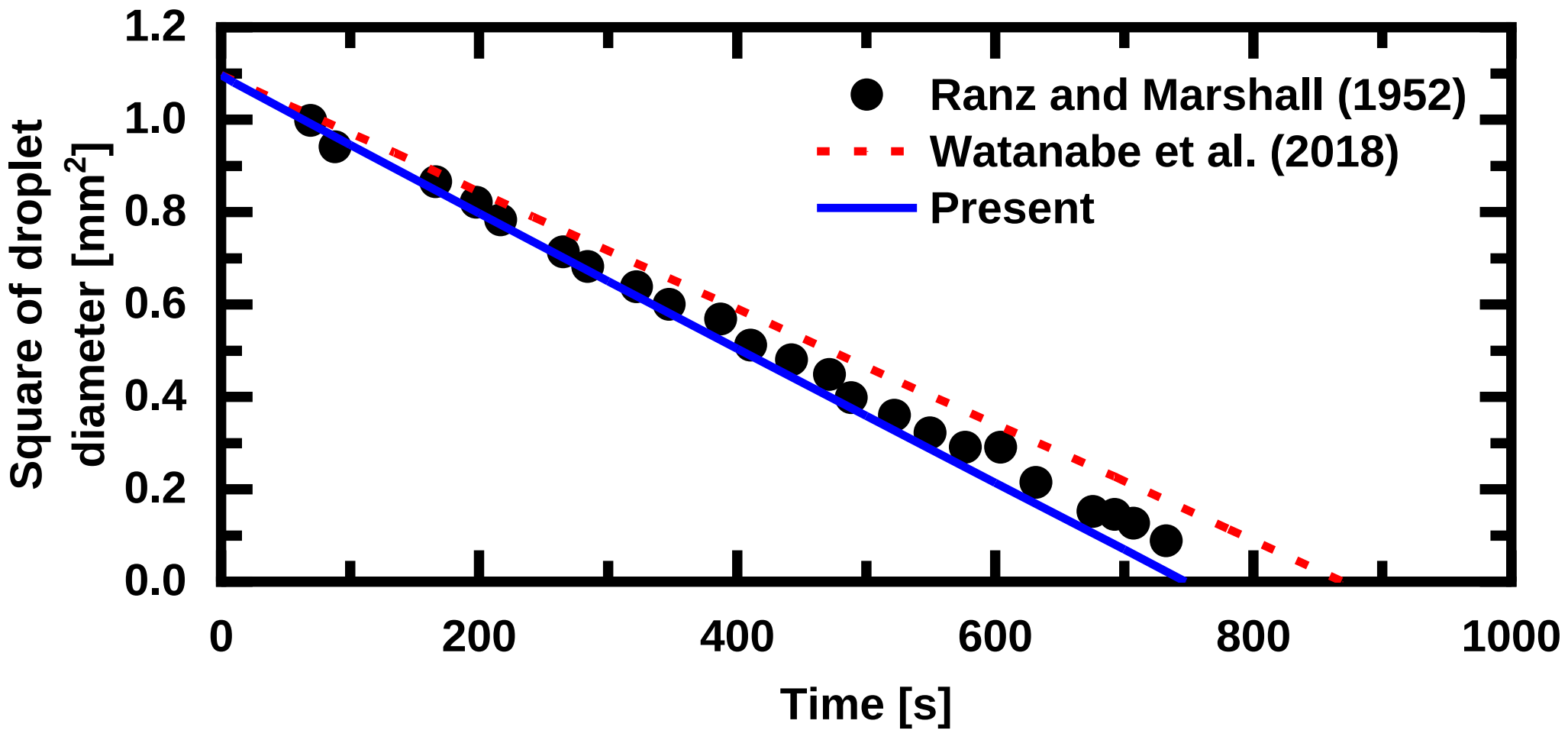

Figure 11. Time history of the diameter of an evaporating water droplet.

The droplet evaporation model is further validated against the experimental data of an evaporating water droplet presented by Ranz and Marshall [31]. The initial droplet diameter and temperature are 1.047 mm and 282 K, respectively, and the surrounding gas temperature is 298 K [31]. It can be seen from Fig. 11 that the evaporation rate coefficient (the slope of $d^2 \sim t$ curve) is slightly over-estimated (by about 4.2%). However, the time history of the droplet diameter predicted by RYrhoCentralFoam shows satisfactory agreement with the experimental data [31]. Note that there are always some uncertainties (e.g. mixed heat transfer modes and perturbed ambient flow environment) in single droplet evaporation experiments, which cannot be accurately quantified or considered in the simulations [18]. Meanwhile, this accuracy of the evaporation model in RYrhoCentralFoam is similar to that (under-predicted by about 10.4%) of the work by Watanabe et al. [65], where Abramzon and Sirignano model [29] is employed.

### *3.3.2 Drag force model*

The drag force model for droplet momentum equation, i.e. Eq. (11), is verified through reproducing the velocity evolutions of initially stationary droplet in a flowing gas. The corresponding droplet velocity evolution can also be obtained through integrating Eq. (11) assuming constant momentum response time. This assumption is valid when $Re_d$ << 1 and droplet evaporation is negligible. It reads

$$\mathbf{u}_d(t) = \mathbf{u} - [\mathbf{u} - \mathbf{u}_d(t_0)] \cdot exp\left(\frac{-t}{\tau_{mom}}\right). \tag{46}$$

Here $\tau_{mom}$ is the momentum response timescale, i.e. [7]

$$\tau_{mom} = \frac{\rho_d d_d^2}{18\mu}. \tag{47}$$

Drag-induced momentum transfer between a non-evaporating droplet and air stream with constant velocity is simulated in a 1-m-long duct. The initial temperature and velocity of the droplet are 300 K and 0 m/s, respectively. Those of the air are 300 K and 10 m/s, respectively. Three momentum response times are chosen, i.e. $\tau_{mom}$ = 0.03, 0.12, and 0.27 s, which respectively correspond to droplet diameters of 100, 200, and 300 μm.

Figure 12 shows the evolutions of droplet velocity at different momentum response times. Excellent agreement is found between the present simulations and the analytical solutions (i.e. Eq. 46) for all the cases, indicating that the drag force model is correctly implemented in RYrhoCentralFoam solver.

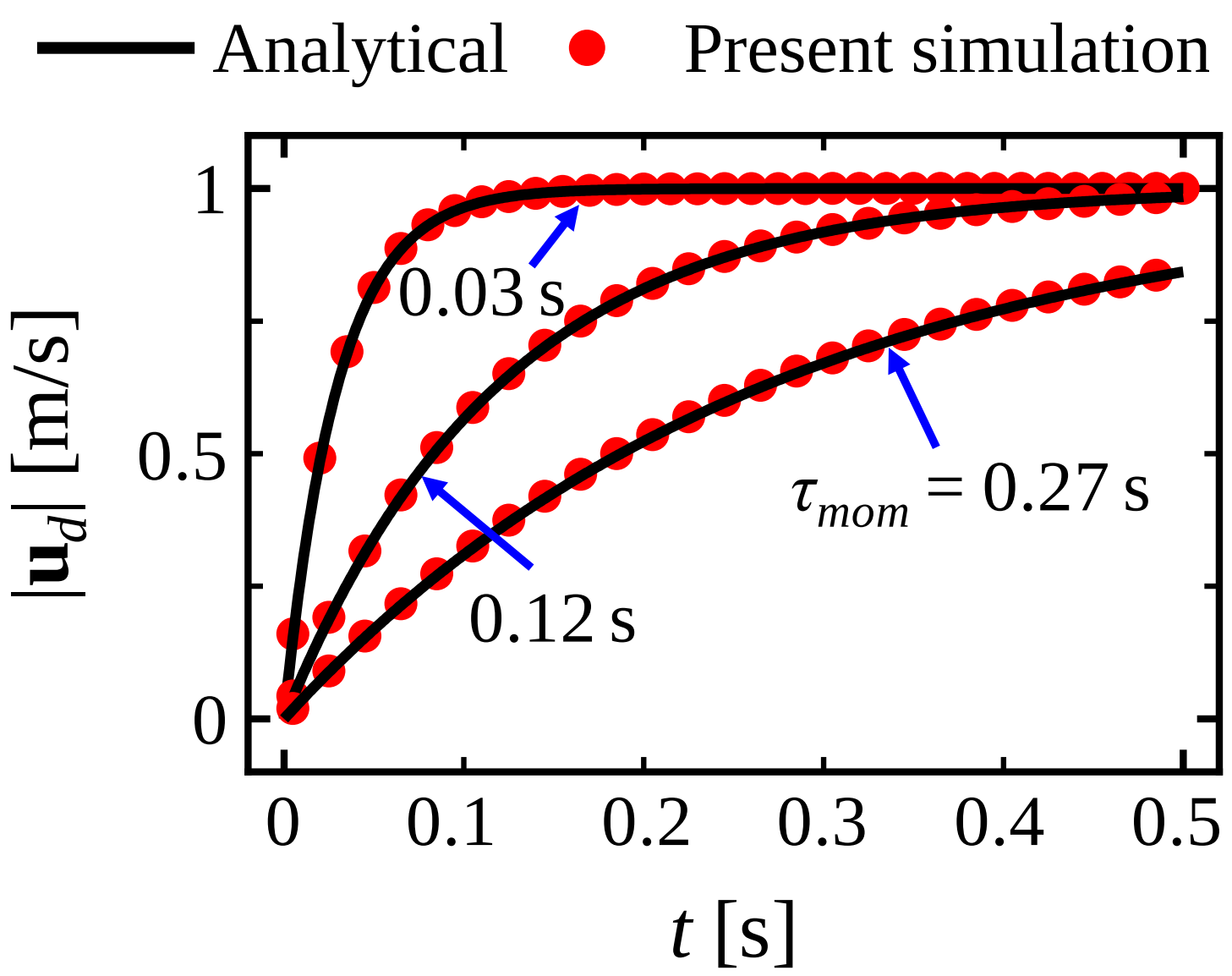

Figure 12. Time history of droplet velocity under different momentum response times.

*3.3.3 Convective heat transfer model*

The convective heat transfer model based on Ranz and Marshall correlation [31] is verified through simulating the heat transfer between quiescent droplet and air. The evolutions of the droplet temperature can also be obtained by integrating Eq. (12) assuming constant thermal response time. This is valid when there is no evaporation and $Re_d$ << 1. For constant temperature of the gas phase, one has the following for the droplet temperature

$$T_d(t) = T - [T - T_d(t_0)] \cdot exp\left(\frac{-t}{\tau_{thermo}}\right). \tag{48}$$

Here $\tau_{thermo}$ is the thermal response timescale, i.e. [7]

$$\tau_{thermo} = \frac{c_{p,d}\rho_d d_d^2}{6Nuk}. \tag{49}$$

The air temperature and velocity are 300 K and 0 m/s, respectively. Those of droplet phase are 400 K

and 0 m/s, respectively. The Nusselt number is 2.0 according to Eq. (26). Three thermal response times of droplet are chosen, i.e. 1.0, 0.6, and 0.2 s.

Figure 13 shows the evolution of droplet temperature at different thermal response times. The results from the RYrhoCentralFoam solver agree very well with the analytical solutions for all the three cases. Only slight difference is found when time increases, probably due to assumption of constant thermal response time is not strictly true in the simulations. Indeed, both droplet density and heat capacity change with droplet temperature, as described by Eqs. (13) and (14), respectively. However, generally, the comparisons in Fig. 13 verify the implementations of convective heat transfer model in our solver.

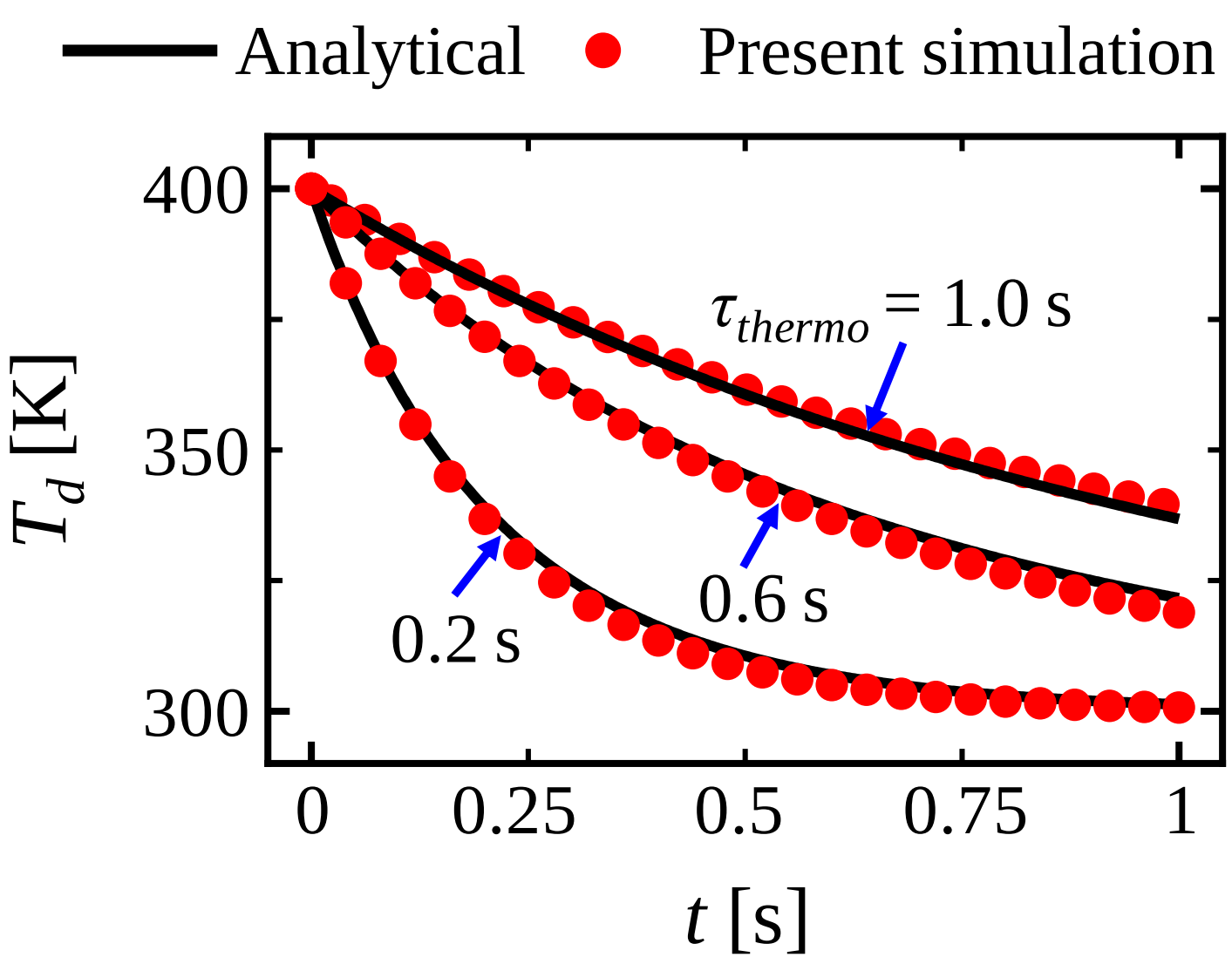

Figure 13. Time history of droplet temperature under different thermal response times.

*3.3.4 Coupling between droplet and gas phases*

In the foregoing sub-sections, the implementations of individual droplet sub-models are verified and/or validated. Here, the interphase coupling in terms of mass, momentum and energy is further validated. To this end, 1D simulations of droplet-laden flows are performed. Water droplets are injected into a 6.096 m long duct, filled with wet air (0.3175% of $H_2O$ vapor in mass fraction). The initial temperature, velocity, and density of droplets are 333.33 K, 30.48 m/s, 1,000 kg/m$^3$, respectively. The gas temperature is 273.33 K with a uniform velocity of 67.056 m/s. Three droplet diameters are studied, i.e. 25, 100, and 1,000 μm. These conditions are identical to those in the analytical solutions by Willbanks et al. [66] and computed ones by Kersey et al. [64] for comparisons.

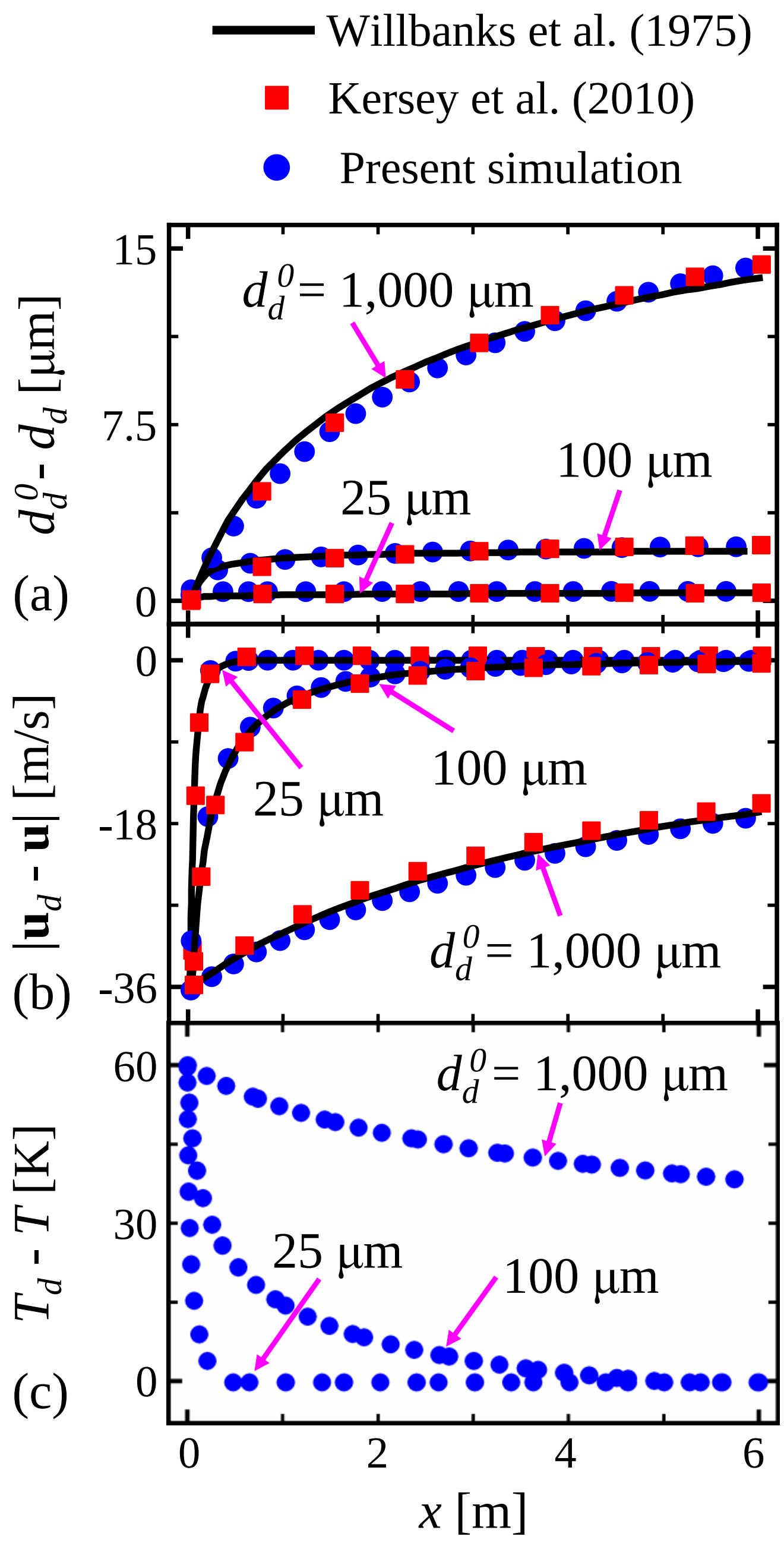

Figure 14. Distributions of (a) droplet diameter decrease, (b) interphase velocity difference, and (c) interphase temperature difference.

Figure 14 shows the variations in droplet diameter, $d_d(t_0) - d_d(x)$, the relative velocity, $|\mathbf{u}_d - \mathbf{u}|$, and

the temperature difference, $T_d$ - $T$, as functions of droplet axial location. Results are compared with the analytical solutions of Willbanks et al. [66] and numerical results of Kersey et al. [64]. Good agreement is found between our numerical results and theirs for all the three cases, for both droplet evaporation and momentum exchange. However, the temperature evolution data are not available in Refs. [62,66] for comparison. The tendencies of temperature evolution in Fig. 14(c) are reasonable for the three cases with difference droplet diameters. In general, based on Fig. 14, two-phase coupling is accurately predicted with RYrhoCentralFoam solver.

### *3.4 Two-phase detonation*

#### *3.4.1 One-dimensional two-phase n-hexane/air or n-hexane/oxygen detonation*

In this sub-section, the accuracy of the RYrhoCentralFoam solver in calculating the detonation propagation speed in two-phase mixtures is studied. One-dimensional two-phase planar detonations in *n*-hexane/air or *n*-hexane/oxygen mixtures are simulated, and various liquid equivalence ratios and droplet diameters are considered. Here the liquid equivalence ratio is defined as the mass ratio of the liquid fuel to the oxidant, scaled by the fuel/oxidant mass ratio under stoichiometric condition. The length of the 1D domain here is 1 m and the uniform mesh size of 0.1 mm is used. It is acknowledged that this mesh resolution does not resolve the induction length. Nonetheless, the sufficiency of the current mesh for calculations of detonation propagation speed has been further checked through mesh sensitivity analysis. The results (not included here) show that the current mesh (0.1 mm) and a finer one (0.01 mm) give close detonation propagation speeds (1,843 m/s and 1,857 m/s, respectively) for the two-phase *n*-hexane/air mixture with equivalence ratio of 1.0. For the *n*-hexane/oxygen mixture, the liquid *n*-hexane equivalence ratio ranges from 0.41 to 0.68 with droplet diameter of 50 μm. For the *n*-hexane/air system, the liquid fuel droplet equivalence ratio is 1.0 with the droplet diameter of 5 μm. The droplet volume fractions are $8.7 \times 10^{-5}$ and $1.5 \times 10^{-4}$ for *n*-hexane/air and *n*-hexane/oxygen mixtures, respectively. Note that no pre-vaporization is considered in our simulations. The initial gas temperature and pressure are 300 K and 1 atm respectively, while the initial droplet temperature is 300

K. One-step mechanism (including 5 species, i.e. *n*-$C_6H_{14}$, $O_2$, $H_2O$, $CO_2$ and $N_2$) [67] is used for *n*-hexane combustion. Its accuracy in detonation simulations has been validated with a skeletal mechanism [68] (See Appendix A).

Figure 15 shows the detonation propagation speed in gas−droplet two-phase mixtures under different conditions. The present results from the RYrhoCentralFoam solver are compared with the experimental data [69,70]. Here the C−J speeds [71] are also added for comparisons, which correspond to the premixtures with fully vaporized liquid fuels. It is shown that the present predicted detonation propagation speed at different conditions is very close to that measured in the experiments (maximum error of 8.2% when liquid equivalence ratio is 0.41). However, they are much less than the C−J speeds of the corresponding purely gaseous mixture. This may be caused by the droplet evaporation and vapor mixing with the surrounding oxidizer. In general, the RYrhoCentralFoam solver and numerical methods can satisfactorily predict the 1D two-phase detonation propagation speed.

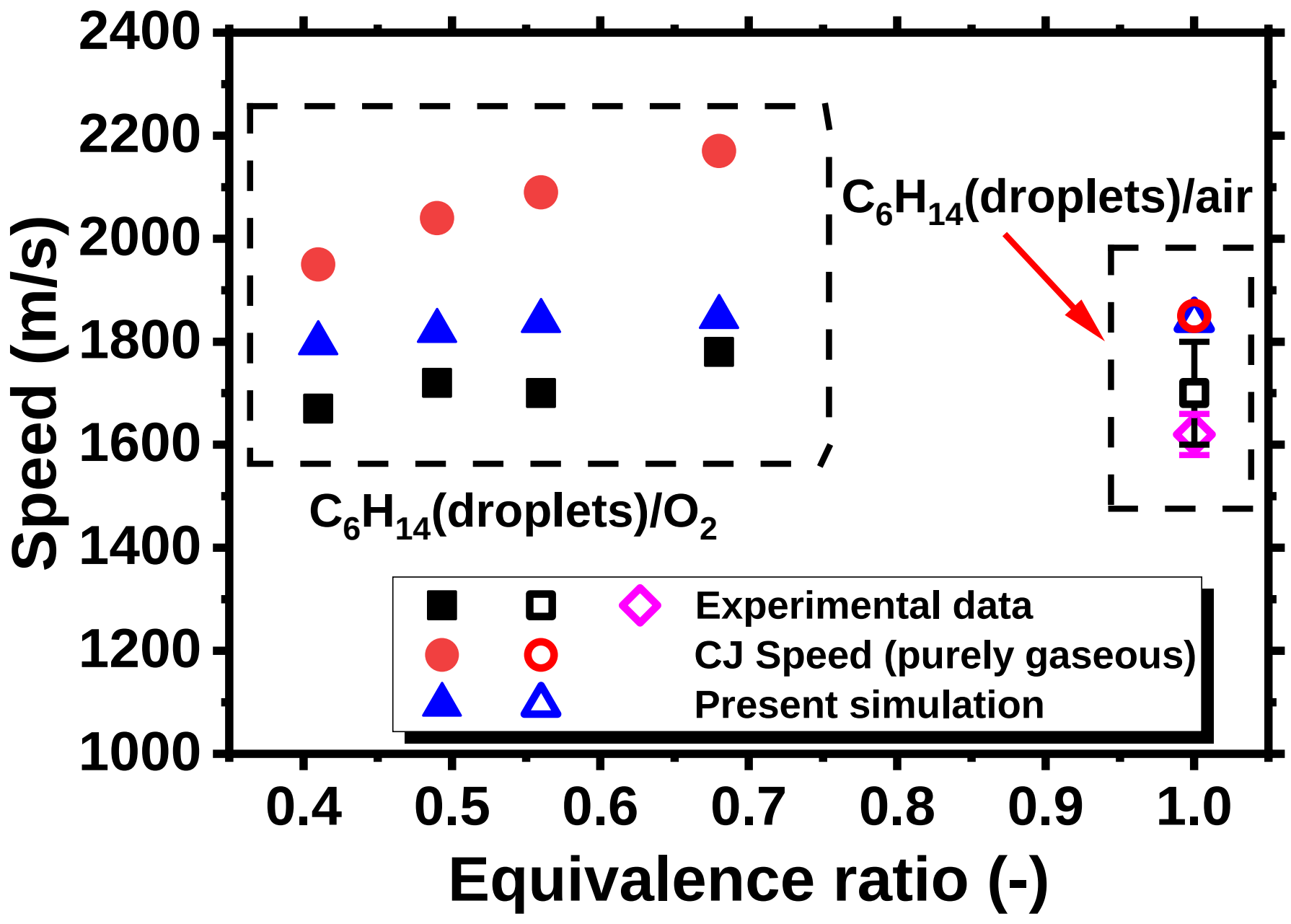

Figure 15. Gas-droplet detonation propagation speed at different conditions. Solid symbol: *n*-hexane/oxygen mixture. Open symbol: *n*-hexane/air mixture. The experimental data are from Refs. [69,70], whilst the C−J speeds are from Ref. [71].

*3.4.2 Two-dimensional detonation in water-droplet-laden ethylene/air mixtures*

Two-dimensional detonation in stoichiometric $C_2H_4$/air gas with water droplets are simulated to examine the capacity of RYrhoCentralFoam solver in predicting interphase coupling and detonation front cellular structure in two-phase mixture with non-reacting sprays. Similar strategy for mesh generation to that in Fig. 8 is used. The length of the two-phase section is 0.1 m after a driver section (0.4 m), and the mesh size is 0.05 mm. The initial gas in the domain is stoichiometric $C_2H_4$/air mixture with $T_0$ = 300 K and $P_0$ = 1 atm. The HRL of the detonable mixture is 0.98 mm. Therefore, the resolution corresponds to approximately 19 cells per HRL. The mono-sized water droplets with diameter $d_d^0 =$ 11 μm and temperature of 300 K are distributed uniformly in the two-phase section, and their mass fraction is 7.1%. The initial water droplet volume fraction is $9 \times 10^{-5}$. Besides, a reduced mechanism for $C_2H_4$ combustion with 10 species and 10 elementary reactions is used [72].

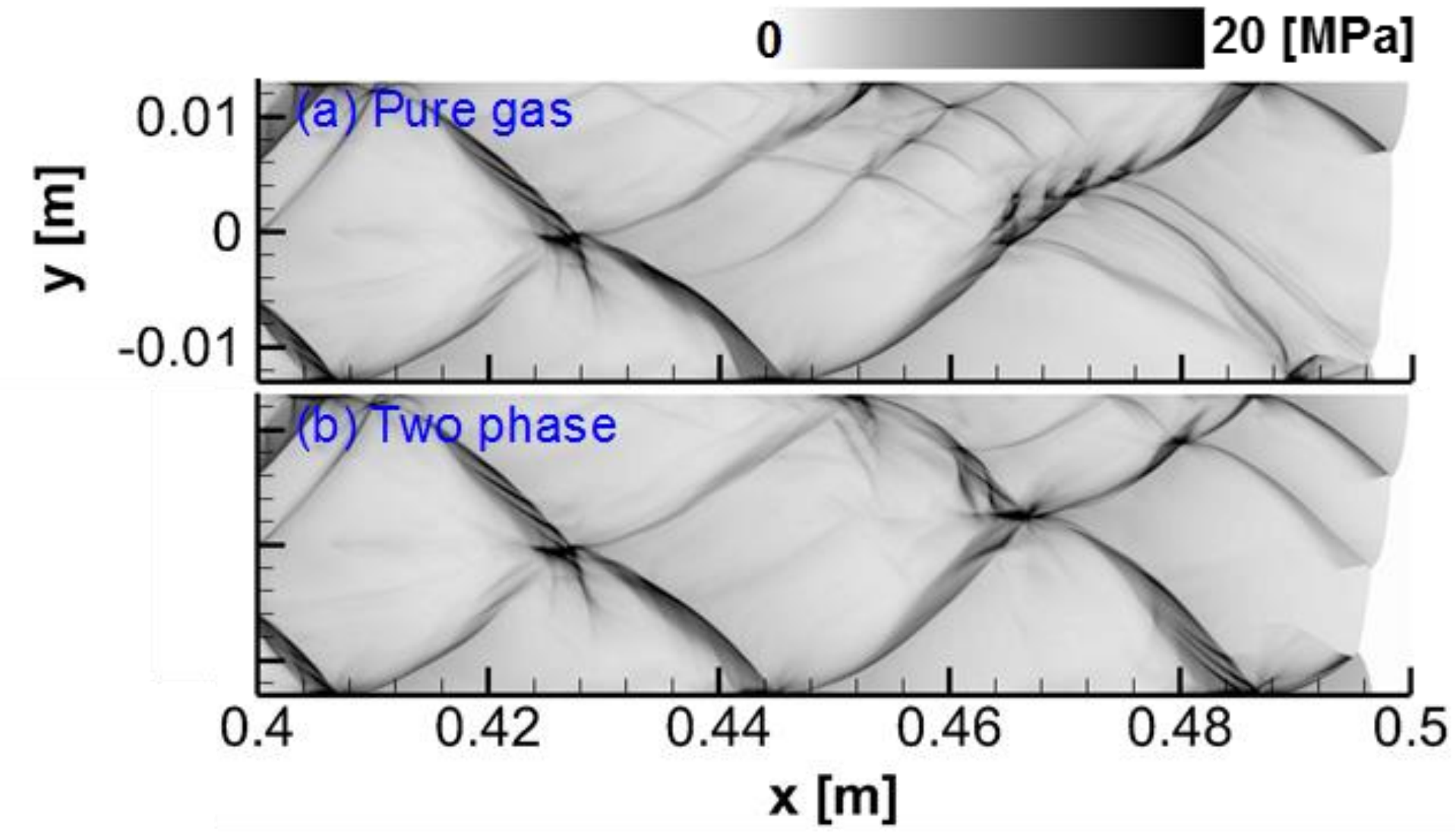

Figure 16. Peak pressure trajectory of detonation wave in (a) pure gas and (b) stoichiometric $C_2H_4$/air mixture with water droplets.

Figure 16 presents the effects of fine water droplets on the detonation cell structure. The cellular pattern in pure gas in Fig. 16(a) is irregular. The detonation wave propagates stably in water spray in Fig. 17(b), and the cell pattern are more regular compared to that of the gaseous detonation. The average cell width of these two cases is approximately 26 mm, which agrees well with the theoretical values [63] and experimental data [73], as tabulated in Table 3.

Table 3. Cell widths of stoichiometric $C_2H_4$/air mixture.

| | | Simulation | Experiment | | Theory |
|---|---|---|---|---|---|
| | | Present work | Bull et al. [74] | Jarsalé et al. [73] | Ng et al. [63] |
| Initial condition ($T_0$, $P_0$) | | 300 K, 100 kPa | 300 K, 100 kPa | 300 K,100 kPa | 300 K, 100 kPa |
| Cell width λ [mm] | Pure gas | ~26 mm | 24.3 mm | 26.5 mm ($\phi$ = 1.02)† | 27.6 mm |
| | Two-phase | ~25.4 mm | – | 42.9 mm ($\phi$ = 1.02) | 39.3 mm |

† The equivalence ratio in the brackets indicate the actual value in the experiments.

The effects of water droplets on the gaseous detonation wave are analyzed in Fig. 17. The strong unstable detonation wave is observed in Fig. 17(a), as indicated by gas temperature. No unburned gas pockets are formed in the downstream of the leading shock front. Basic detonation frontal structures, e.g. Mach stem, incident wave, transverse wave, primary triple point, and secondary triple point, are identified in Fig. 17(b). Figure 17(c) shows that chemical reactions mainly appear behind the leading shock front. In Figs. 17(d)−(f), the presence of water droplets changes the two-phase detonation flow fields significantly. An egg-shaped structure, which is composed of transverse waves and reflection waves, is formed behind the Mach stem.

It can also be observed in Fig. 17(g) that the water droplets experience a finite distance to get heated towards its saturated temperature and the relaxation distance is about 2 mm before the saturated temperature is reached. Large evaporation rate in Fig. 17(h) occurs behind the Mach stem and the upper portion of leading front, which corresponds to high heat release rate in Fig. 17(f). Combining Figs. 17(d)−(i), we can see that within relatively large denoted area, water droplet vaporization is not completed, and hence the continuous interactions between the liquid and gas phases can be expected.

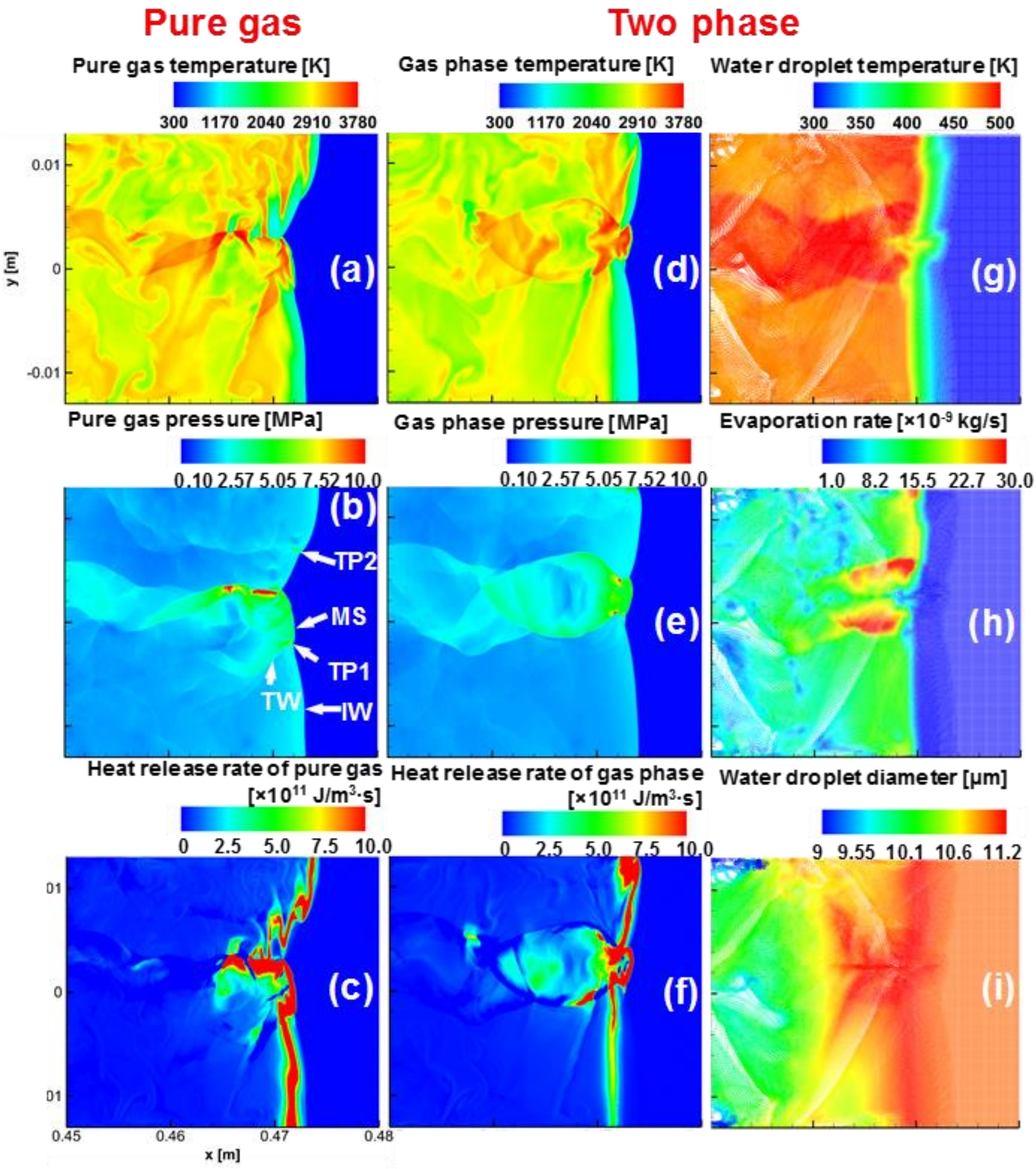

Figure 17. Pure gas detonation: (a) temperature, (b) pressure and (c) heat release rate. Two-phase detonation: (d) gas temperature, (e) gas pressure, (f) heat release rate, (g) Lagrangian water droplets colored with droplet temperature, (h) evaporation rate and (i) droplet diameter. The detonation wave propagates from left to right side. MS: Mach stem, TP1: primary triple point, TP2: secondary triple point, IW: incident wave, TW: transverse wave.

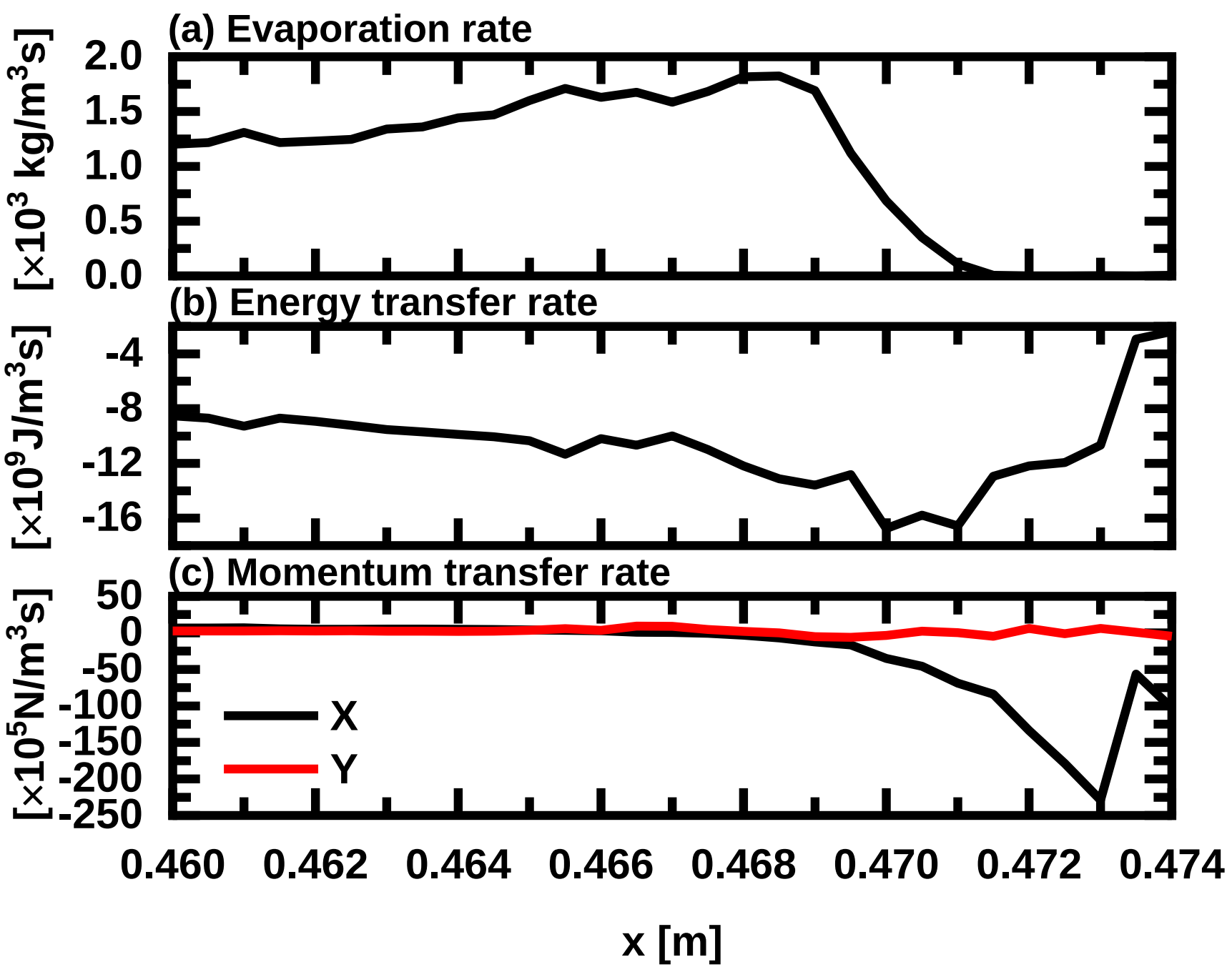

Figure 18. Width-averaged (a) evaporation rate, (b) energy transfer rate and (c) momentum transfer rate along $x$- and $y$-directions. The leading shock front is located at $x$ = 0.474 m.

The width-averaged interphase exchange rates calculated with Eqs. (28)–(30) are presented in Fig. 18, which corresponds to the same instant in Fig. 17. It is observed in Fig. 18(a) that evaporation rate is suppressed immediately behind the detonation wave ($x$ = 0.471−0.474 m), and peaks at $x$ = 0.468−0.469 m. This is caused by the elevated pressure behind the leading shock front and increased water vapour concentration due to the chemical reactions. Moreover, the energy transfer rate in Fig. 18(b) increases within the suppression region, and then decreases slightly with recovered evaporation rate. This is because the energy exchange between the gas phase and liquid droplets is promoted by the chemical reaction which mainly occurs behind the leading shock front and part of the transverse detonation. However, in the downstream of the detonation wave the low reaction rate weakens energy exchange. In Fig. 18(c) large momentum transfer rate is found in $x$-direction, especially behind the detonation wave, whilst smaller fluctuation of momentum exchange along $y$-direction is seen. This is due to the detonation wave mainly sweeps along the $x$-direction. This case has demonstrated the good prediction abilities of the RYrhoCentralFoam solver for two-phase detonative combustion in fine water sprays.

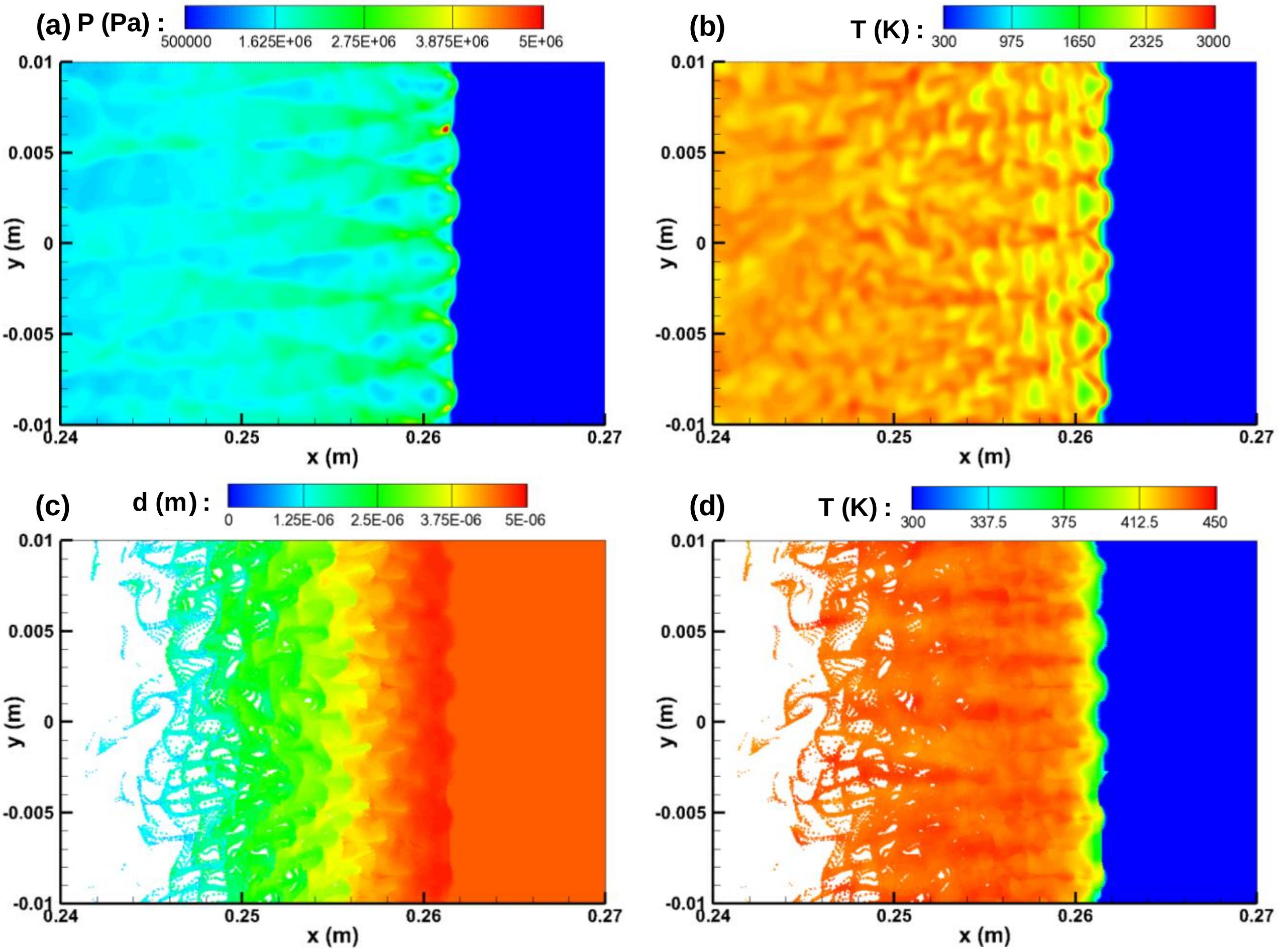

Figure 19. Distributions of (a) pressure, (b) gas temperature, and Lagrangian fuel droplets colored with (c) diameter and (d) temperature.

### *3.4.3 Two-dimensional detonation in two-phase n-hexane/air mixtures*

Numerical simulation of two-dimensional detonation in two-phase *n*-$C_6H_{14}$/air mixture is conducted in this sub-section. Here the length and height of the computational domain are 0.3 m and 0.02 m, respectively. Zero gradient condition is enforced for the left and right sides, whilst slip wall conditions are assumed for the upper and lower boundaries. The uniform mesh size of 0.05 mm is used. The two-phase *n*-$C_6H_{14}$/air mixtures include *n*-$C_6H_{14}$ vapor and liquid *n*-$C_6H_{14}$ droplets with uniform diameters of 5 $\mu$m. The respective equivalence ratios of vapor and droplets are 0.5, corresponding to the total equivalence ratio is 1.0. The initial gas temperature and pressure are set as 300 K and 1 atm, respectively. The initial temperature of the droplets is 300 K, and the initial volume fraction is 0.00015. For *n*-hexane/air combustion, one-step mechanism (including 5 species, i.e. *n*-$C_6H_{14}$, $O_2$, $H_2O$, $CO_2$ and $N_2$) [67] is used, which is also used in section 3.4.1.

Figure 19 shows the distributions of gas pressure and temperature, as well as the Lagrangian *n*-hexane droplets colored with droplet diameter and temperature. As shown in Figs. 19(a) and 19(b), the detonation propagates stably in the two-phase *n*-$C_6H_{14}$/air mixtures and the basic detonation structures such as the Mach stem, incident shock wave, transverse wave and triple point are captured. Stripe structures of gas temperature (see Fig. 19b) are also observed behind the detonation front, which may be due to the interactions between the Mach stem, incident shock wave and the fuel droplets. The effects of the basic detonation structures on the fuel droplets can be observed with the distributions of droplets diameters and temperature as shown in Figs. 19(c) and 19(d). It can be seen that the droplets exist for a distance of about 20 mm behind the detonation front before they are evaporated completely, where the vapor from the droplet would in turn affect the detonation structures and the detonation propagation. The fuel droplets experience a distance of about 2 mm to get heated towards its saturated temperature behind the detonation front as shown in Fig. 19(d). The upward or downward movement of transverse waves leads to the irregular distributions of the droplets, which makes the temperature distributions behind the detonation front ($x < 0.255$ m) "turbulent" (see Fig. 19b). Moreover, it should be noted that the mesh resolution of this case is 0.05 mm, which may be not fine enough to capture the fine structures such as the jet shear layers in detonation propagation. However, the results in this sub-section and Section 3.4.2 have confirmed that the RYrhoCentralFoam solver can be used to simulate the two-dimensional detonation in gas−droplet two-phase mixtures.

## 4. Conclusion

In this work, a gas−droplet two-phase compressible flow solver, RYrhoCentralFoam, is developed based on hybrid Eulerian-Lagrangian method to simulate the two-phase detonative combustion. For Eulerian gas phase, RYrhoCentralFoam is second order of accuracy in time and space discretizations and based on finite-volume method on polyhedral cells. The following developments are made within the framework of the compressible flow solver rhoCentralFoam in OpenFOAM® [15]: (1) multi-component species transport, (2) detailed fuel chemistry for gas phase combustion, (3) Lagrangian solver for gas−droplet two-phase flows and sub-models for liquid droplets. To verify and validate the developments and implementations of the solver and sub-models, well-chosen benchmark test cases are studied, including non-reacting multi-component single-phase flows, purely gaseous detonations, and two-phase gas-droplet mixtures.

The results show that the RYrhoCentralFoam solver can accurately predict the flow discontinuities (e.g. shock wave and expansion wave), molecular diffusion, auto-ignition as well as shock-induced ignition. Also, the RYrhoCentralFoam solver can accurately simulate detonation propagation for different fuels (e.g. hydrogen and methane), in terms of propagation speed, detailed detonation structure and cell size. Sub-models related to the droplet phase are verified and/or validated against the analytical and/or experimental data. It is found that the RYrhoCentralFoam solver is able to calculate the main features of the gas−droplet two-phase detonations, including detonation propagation speed, interphase interactions and detonation frontal structures.

Moreover, due to the excellent modularization characteristics of OpenFOAM®, the prediction abilities of RYrhoCentralFoam solver can be potentially extended for simulating detonations in dense droplets through introducing the relevant modules, e.g. droplet break-up and collision. This offers an interesting direction for our future investigations.

## Acknowledgements

This work is supported by Singapore Ministry of Education Tier 1 grant (R-265-000-653-114). The computations for this article were fully performed on resources of the National Supercomputing Centre, Singapore (https://www.nscc.sg). Ruixuan Zhu is thanked for calculating the Chapman–Jouguet speeds with SD Toolbox in Fig. 7. Qingyang Meng is acknowledged for sharing the OpenFOAM routines for data post-processing. Professor Hai Wang from Stanford University is thanked for helpful discussion about the JetSURF 2.0 mechanism.

## Appendix A: comparison of *n*-hexane chemical mechanism

The one-step chemistry of *n*-hexane for detonation combustion used in Sections 3.4.1 and 3.4.3 is validated with a skeletal mechanism (JetSurF 2.0) [68]. It is found from Fig. A1 that the results from the one-step mechanism [67] show good agreement with the those from the skeletal mechanism [68], except the equivalence ratio close to 1.0. In general, the one-step mechanism is accurate for predictions of the key parameters in *n*-hexane/air detonation.

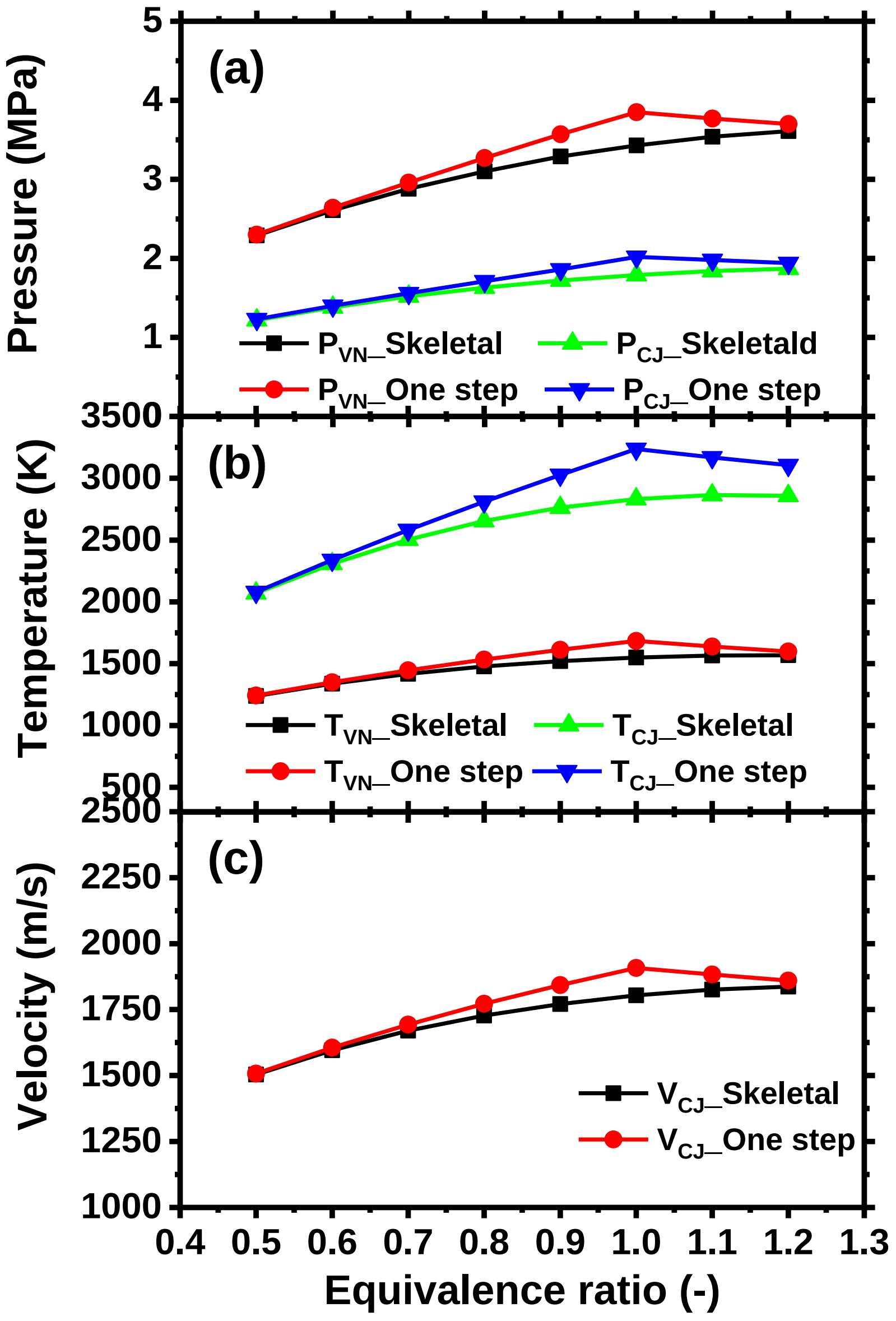

Figure A1. Comparisons between one-step [67] and skeletal mechanisms [68] for *n*-hexane/air mixture: (a) ZND and C-J pressure, (b) ZND and C-J temperature and (c) C-J velocity.